\documentclass[12pt]{article}

\usepackage{epsfig,amsfonts,newlfont,rotate,float,amssymb}

\DeclareMathAlphabet{\mathitbf}{T1}{cmr}{bx}{it}

\newcommand{\e}{\mathrm e}
\newcommand{\openone}{{\mbox{\boldmath $1$}}}
\begin{document}
\title{
High-$T_c$ superconductivity through a charge pairing
mechanism in a strongly coupled disordered phase.}

\author{
J.L.~Alonso$^1$, Ph.\ Boucaud$^2$, \\
V.~Mart\'{\i}n-Mayor$^3$ and A.J.~van der Sijs$^4$
}

\date{April 13, 1998}
\maketitle
\thispagestyle{empty}

\noindent {\small {\it $^1$Departamento de F\'\i sica Te\'orica, Universidad de Zaragoza,
50009 Zaragoza, Spain.}}

\noindent {\small {\it $^2$LPTHE, Universit\'e de Paris XI,
91405 Orsay Cedex, France.}}

\noindent {\small {\it $^3$Departamento de F\'{\i}sica Te\'orica I,
Universidad Complutense de Madrid, 28040 Madrid, Spain.}}

\noindent {\small {\it $^4$Swiss Center for Scientific Computing, ETH-Z\"urich,
ETH-Zentrum, CH-8092 Z\"urich, Switzerland.}

\noindent {\small {\bf e-mail:} {\it 
$^1$buj@gteorico.unizar.es, $^2$phi@qcd.th.u-psud.fr,$^3$victor@lattice.fis.ucm.es,}}

\noindent {\small {\it $^4$arjan@scsc.ethz.ch}}

\begin{abstract}
We present a lattice field theory of spins coupled to
Dirac fermions, as a model for the doped copper oxide compounds.
Both the fermionic and spin degrees of freedom
are treated dynamically. The influence of the charge carriers on the
magnetic
ordering follows automatically.
The magnetic phase diagram at zero temperature is studied 
numerically and with Mean-Field methods. The Hybrid  Monte Carlo 
algorithm is adapted to
the O(3) non-linear $\sigma$ model constraint.
The  charged excitations in the various phases 
are studied at the Mean-Field level.
Bound states of two charged fermions are found in a strongly coupled 
{\it paramagnetic\/} phase, without requiring a Fermi sea.
We acquire a qualitative understanding of high-T$_{\mathrm c}$ 
superconductivity through
Bose-Einstein condensation of those dynamically bound pairs.
The model also implies insulating  behaviour
at low doping, and Fermi liquid behaviour at large doping.
We predict the possibility of
reentrant superconductivity and the absence of superconductivity in 
spin-$1$ and spin-$3/2$ materials.
\end{abstract}

\vskip 5 mm

\medskip


\vskip 5 mm

\vglue -22.8cm
\hfill DFTUZ/98/12, LPTHE-ORSAY 98/23, SCSC-TR-98-03

\newpage

%123456789%123456789%123456789%123456789%123456789%123456789%123456789%1
\section{Introduction}

\label{intro}

Since the discovery of the Cu-based high-$T_{\mathrm c}$ superconductors \cite{BM}
there has been a lot of interest in
the physics of antiferromagnets doped with holes.
The understanding of such a system would allow a unifying
picture of the magnetic and superconducting properties of those materials.
Some of the properties to be considered are:
\begin{enumerate}

\item Superconductivity appears in doped, ceramic materials, like La$_2$CuO$_4$
doped to La$_{2-x}$Sr$_x$CuO$_4$, {\it in a narrow range of doping not too
far from the antiferromagnetic insulating state}, at low temperatures.

\item For zero doping fraction, $x=0$, they are insulating antiferromagnets.

\item At large $x$ they appear to have normal metallic behaviour.

\item They are layered compounds, made up of CuO$_2$ planes separated
by some ``charge reservoir''. 
The doping suppresses the antiferromagnetic (AFM) correlation between
neighboring CuO$_2$ planes. However, large ($\xi\sim 10a$) in-plane 
correlations remain just above $T_{\mathrm c}$. Also, transport phenomena occur 
mainly in the CuO$_2$ planes, so everything looks like a $d$=2 problem, 
{\it with localized spins on the Cu and mobile holes on the O ions}.

\item As in the conventional superconductors, the superconductivity
is charge-2 in nature, suggesting a pairing state.

\item The {\it normal} ({\em i.e.} non superconducting) state is anomalous:
there is experimental evidence for anomalous behaviour of several
response functions (magnetic susceptibility, specific-heat, DC conductivity)
in the {\it normal} phase. This is usually refered to as a 
{\it pseudo-gap} phase, for which a picture of a singlet pairing already
{\it above} $T_{\mathrm c}$ is emerging \cite{RANDERIA}. That is, the pair formation 
seems to occur {\it independently} of the quantum liquid condensation
({\em i.e.} the macroscopic phase coherence), at least for underdoped 
({\em i.e.} superconducting samples less than optimally doped) 
materials~\cite{PARES}.

\item There is a negligibly small isotope effect and there is evidence
\cite{TSUEI} for a $d_{x^2-y^2}$ wave pairing state.

\end{enumerate}

The essential questions posed by the phenomenology
of the new perovskite superconductors have been phrased as follows
in Ref.~\cite{PINES}:
\begin{itemize}
\item
What is the physical origin of the anomalous {\it normal} state?
\item
How can it be characterized?
\item
What is the mechanism for high-$T_{\mathrm c}$ superconductivity?
\item
What is the pairing state?
\end{itemize}
We will try to formulate an answer to these questions in the discussion
section of this paper.

The behaviour of the undoped compounds (point 2.\ above) is well
understood in terms of the
two-dimensional nearest-neighbour quantum antiferromagnetic (AFM)
Heisenberg model.
This model can be mapped to the O(3)
non-linear $\sigma$-model~\cite{haldane},
and the predictions, obtained
through renormalization group analysis~\cite{CHAKRA} and chiral perturbation
theory~\cite{HASENFRATZ}, agree very well both with
experimental~\cite{ENDOH} data for the cuprates and with numerical
\cite{DING} data from direct simulation of the AFM Heisenberg model.
Therefore, it seems natural to take either the AFM Heisenberg model or
the O(3) sigma model as a starting point to study the doped materials.
In contrast with the successful description of the undoped compounds,
however, to our knowledge no model has yet been capable of describing
the full doping range. 

We have recently proposed a simple model~\cite{LETTER,HEPLAT} which is
a natural extension of the O(3) model to include doped charge carriers.
The model is able to explain, at least
qualitatively, many of the observed properties of the perovskites from
undoped to (highly) overdoped compounds.  It is a 
lattice-regularized, field-theoretical model of interacting spins and
Dirac fermions in 2+1 dimensions, with only two free parameters in addition
to the temperature: a nearest-neighbour spin coupling and a
spin-fermion coupling.

Conceptually, the model has much in common with
microscopic spin-fermion Hamiltonians~\cite{MURAMATSU1}, which are
somewhat less restrictive than the $t$-$J$ model~\cite{ZHANG}.
The formulation of our model as a local lattice field theory
has several advantages over those and other models, though. Mean-field
(MF) calculations are feasible and, in the absence of a sign problem,
numerical Monte Carlo (MC) simulations can be carried out. Our MC
results furthermore demonstrate that the MF approximation is quite
reliable here.
 
In the present article we want to present a careful, detailed discussion
of the model, its symmetries, and its properties,
and give full technical details and results of the MF and MC calculations,
some of which were reported in Ref.\ \cite{LETTER}. 
In the present paper, our mean-field and numerical studies
will be limited to the zero-temperature case, 
corresponding to infinite Euclidean time direction.
However, we will argue on general grounds what
happens when the temperature is increased.

The remainder of this paper is laid out as follows. In Section 2 we
present our model.
In Section \ref{SYM} we discuss the choice of lattice fermions, comment on
the symmetries of the model, give its phase diagram and
prove the reality of the fermion determinant, even in the
presence of a chemical potential.  In Section 2.2, we give
our interpretation of the model as an effective
field theory which tries to embody the essential features of holes
strongly coupled to a dynamical AFM spin background, the only
fundamental ingredients being symmetries. 
In Section 3 we examine the phase diagram of the model in the MF approximation.
In Section 4 we use MC simulations to complete the study of the phase diagram.
For this purpose we have developed a new method that exactly solves the technical
problem related to the length-1 constraint on the spin variables \cite{PARISI}.
Section 5 is devoted to a study of the relevant excitations in the
different phases of the system, at the MF level.
A crucial result is the dynamical generation of spin singlet bosonic
bound states of charged fermions in the so-called
paramagnetic strong (PMS) phase. At the MF level we
have not detected any light {\em fermionic\/} excitations at {\it zero}
temperature in this PMS phase.
Another interesting result is the emergence
of fermionic excitations around spatial momenta
$(\pm\pi/2,\pm\pi/2)$~\cite{CHUBUKOV,MARSHALL}, in the low doping AFM 
phase preceding the superconducting (SC) phase.  An important
consequence of the paramagnetic nature of our SC phase is that
fermionic excitations, if any, {\it will not
center around} $(\pm\pi/2,\pm\pi/2)$, while 
such a behaviour is typical
of an AFM phase (see ref.~\cite{NORMAN} for some recent experimental
results).  Next, in Section 6.1, we analyze how our model describes
the various phenomena in the cuprates at zero temperature and in
Section 6.2 we conjecture the qualitative behaviour of the model at non
zero temperature, using arguments based on a few very general
thermodynamical properties.

In the Discussion section (Section 7) we will try to answer the
essential questions pointed out in Ref.\ \cite{PINES} and quoted above.
We also reconsider points 1 to 7 as well as the Bose-Einstein (BE) to BCS
crossover from the perspective of our model,
and we comment on our predictions.

\section{The model}

\subsection{Motivation, Formulation, Symmetries, Phase Diagram}

\label{SYM}

In order to motivate our model, we start by collecting some of
the most compelling pieces of experimental and theoretical evidence
and turning them into guidelines for the formulation of the model.

First of all, recall that the undoped parent compounds are described
excellently
by the 2+1 dimensional O(3) non-linear $\sigma$-model
\cite{CHAKRA,HASENFRATZ,ENDOH,DING} (see also Ref.\ \cite{WIESE97}).
Therefore, it is very natural to take this as our first guideline:

{\bf A. In the limit of zero doping, $x=0$, our model should reduce to
the O(3) non-linear $\sigma$-model}.  

Another important piece of information comes from point 7 of the list
in Section 1.
The virtual absence of an isotope effect and the evidence for
a $d_{x^2-y^2}$ wave pairing state
favor a mechanism based on spin rather than
phonon-mediated pair formation~\cite{SCHRIEFFER,ANDSCHRI}.
This leads us to our second guideline:

{\bf B: Our model should be a spin-fermion model} describing fermions
of varying mobility
interacting with a dynamical background of localized spins.

Next, as emphasized in~\cite{BIRGENEAU}, the general topology of the
cuprates phase diagram was in fact first predicted in Ref.\
\cite{AHARONY}. 
Their point is to assume a strong coupling between the spin of the hole
on the oxygen (see point 4) and the spin of the surrounding Cu$^{2+}$
ions due to the direct overlap of the wavefunctions.
Therefore, from this:

{\bf C: We expect the strong-coupling regime of our 
spin-fermion model to be the
relevant regime} to understand, eventually, the insulating and
superconducting phases of cuprates.

Notice that to in order to arrive at {\bf A} and {\bf B}
we have essentially used only information from points 2, 4 and 7,
while guideline {\bf C} comes from
the more general observation that we deal with a
strong coupling phenomenon.

Given the above considerations, we have proposed~\cite{LETTER}
a model defined by the following $(2+1)$-dimensional lattice 
euclidean (imaginary time) path integral,

\begin{equation}
Z = \int\!D{\mbox{\boldmath $\phi$}} \, D\bar\psi \, D\psi \, \exp(-S)
\label{Z}
\end{equation}
with action,
\begin{equation}
S = -\sum_{x,\mu} k \, {\mbox{\boldmath $\phi$}}_x\cdot {\mbox{\boldmath $\phi$}}
_{x+\hat\mu}+
 \sum_{x,\mu} \frac{\rho}{2} \bar\psi_x \gamma^\mu (\psi_{x+\hat\mu} -
    \psi_{x-\hat\mu}) 
+\sum_{x} \lambda \, \bar\psi_x {\mbox{\boldmath $\phi$}}_x \cdot {\mbox{\boldmath $
\tau$}} \psi_x\label{ACCIONZ}
    \, .
\end{equation}
We use this expression as our starting point, but it should be noted
that the model
{\it
depends only on the ratio $y=\lambda/\rho$, through a change in
the normalization of the fermion field}. In terms of the effective
spin-fermion coupling $y$, we get:

\begin{equation}
\label{ACCION}
S=-\sum_{x,\mu} k \, {\mbox{\boldmath $\phi$}}_x\cdot {\mbox{\boldmath $\phi$}}
_{x+\hat\mu}+
 \sum_{x,\mu} \, \frac{1}{2}\, \bar\psi_x \gamma^\mu (\psi_{x+\hat\mu} -
    \psi_{x-\hat\mu})
+\sum_{x} \, y\,  \, \bar\psi_x {\mbox{\boldmath $\phi$}}_x \cdot {\mbox{\boldmath $
\tau$}} \psi_x
    \, .
\end{equation}

Here $x$ runs over a (2 + 1)-dimensional cubic Euclidean space-time lattice.
We keep in mind, however, that some coherence perpendicular to the CuO$_2$
planes is required to avoid problems with the Hohenberg-Mermin-Wagner
theorem at finite temperature, and for a Bose-Einstein condensate to form.
Some small supplementary couplings in the orthogonal spatial direction
would play this role.

The fields $\psi$
represent the doped  electric charges.
Each $\psi_x$ is a fermionic four-spinor as a shorthand for two flavours of two-component
Dirac spinors (we use flavour to mimic the spin, because there is no
spin in two spatial dimensions).
Each flavour accounts for two components of the four of a
\hbox{$(3+1)$-dimensional} Dirac electron or hole~\cite{AFFLECK}. 
Both flavours are taken in the same irreducible spinor
representation, with $2\times 2$
gamma matrices taken as the Pauli matrices $\sigma^\mu$.
The $4\times 4$ matrices $\gamma^\mu$ in Eq.\ (\ref{ACCION}) have the form

\begin{equation}
\gamma^\mu = \left( \begin{array}{cc} \sigma^\mu & 0 \\ 0 & \sigma^\mu
    \end{array} \right) \ \ \ \ \ \ \ \ \ \ \ \mu = 1,2,3.
\label{gammas}
\end{equation}

The kinetic term for the
fermions is of the nearest-neighbour (hopping) form. Lattice fermions
defined in this way undergo species doubling in the perturbative continuum
limit~\cite{CONTINUUM}. But since all the fermions, the physical one as well
as the doublers, decouple in the continuum limit performed in the region 
of strong spin-fermion coupling \cite{DE}, the
doubling problem is in fact limited to the weakly coupled region. As our
only relevant conclusion in this region, the occurrence of Fermi liquid 
behaviour because of the weak coupling, is not affected by the
doublers, we canuse these ``naive'' fermions
without any problem.
The three-component fields
${\mbox{\boldmath $\phi$}}$ denote the spins located
at the copper ions. They are real scalar bosonic variables, subject to the
constraint ${\mbox{\boldmath $\phi$}}^2=1$, as in the O(3) non-linear
$\sigma$-model. 
Their kinetic term, a nearest-neighbour hopping
interaction, is the field-theoretic equivalent of a Heisenberg
superexchange interaction.  The last term in Eq. (\ref{ACCION}) describes the
interaction between the spins and the Dirac fermions, which is diagonal in Dirac space. The Pauli
matrices $\tau^a$ act in flavour space.

Let us now consider the symmetries of (\ref{ACCION}).
First of all, we have the usual cubic symmetry,
the lattice remnant of (2+1)-dimensional Euclidean ``Lorentz'' symmetry.
Next there is a discrete parity symmetry, which in (2+1) dimensions
is defined as the reflection of one of the spatial axes, say the $x$-axis.
Under this parity symmetry, the fermions can be seen to transform as
\begin{equation}
\psi \rightarrow \sigma_1 \, \psi, \ \ \ \ \ \ \ \ 
\bar\psi \rightarrow - \bar\psi \, \sigma_1
 \, , \label{psipar}
\end{equation}
so the ${\mbox{\boldmath $\phi$}}$ field is a pseudoscalar in this sense.
In addition, the action (\ref{ACCION}) is invariant under an SU(2) ``flavour''
symmetry in which $\psi$ transforms as the fundamental representation and
${\mbox{\boldmath $\phi$}}$ transforms as the adjoint one.
So, our model maintains  after doping the well tested symmetry of the
O(3) non-linear $\sigma$-model describing the undoped material~\cite{ENDOH}.
Note that by requiring the two fermion flavours to have the same Lorentz
structure (that is, by choosing the $\gamma$'s as in (\ref{gammas}))
no fermion mass term is allowed which preserves the above
symmetries~\cite{APPELQUIST}.

There are two more discrete symmetries of our model (\ref{ACCION}),
which will be useful in the MF calculation of the phase-diagram.
The first one is
\begin{equation}
Z(k,y) = Z(k,-y)
 , \label{symmZ1}
\end{equation}
which becomes clear if we make the change of variables
\begin{eqnarray}
\label{symmfields1}
\psi_x \rightarrow \exp\left(i\frac\pi2 \epsilon_x\right) \, \psi_x, \ \ \ \ 
\bar\psi_x \rightarrow \exp\left(i\frac\pi2 \epsilon_x\right) \, \bar\psi_x,
\end{eqnarray}
where
\begin{equation}
\epsilon_n = (-1)^{\sum_\mu x_\mu}
 . \label{epsilonn}
\end{equation}
This implies that $Z(k,y)$ is a function of $y^2$ only and we can restrict
ourselves to $y>0$.

In addition, there is a symmetry
\begin{equation}
Z(k,y) = Z(-k,-iy)
 , \label{symmZ2}
\end{equation}
as can be seen by making the substitutions
\begin{eqnarray}
\label{symmfields2}
\psi_x \rightarrow \exp\left(i\frac\pi4 \epsilon_x\right) \, \psi_x, \ \ \ \ 
\bar\psi_x \rightarrow \exp\left(i\frac\pi4 \epsilon_x\right) \, \bar\psi_x, \ \ \ \ 
{\mbox{\boldmath$\phi$}}_x \rightarrow \epsilon_x \, {\mbox{\boldmath$\phi$}}_x
 . 
\label{SIMSTAGG}
\end{eqnarray}

The latter symmetry implies that the lattice regularization of
the non-linear $\sigma$-model, $y$=0 (or $y$=$\infty$, see Sections
\ref{MFPD}, \ref{MCPD}), is equally valid in a ferromagnetic 
or an antiferromagnetic phase. This will be of fundamental importance
in section \ref{PHEN}.

In order to perform computations in field theoretical models of this type,
one has to integrate out the fermion fields; the reason being that they
are anticommuting Grassmann fields, while both in MF and in MC one needs
to be able to work with a c-number ordering.
This integration leads to a ${\mbox{\boldmath $\phi$}}$-dependent determinant which is called
the fermion determinant.
It is important to know whether this determinant is a real number.
To study this, let us write down the original fermion matrix 
(Latin letters $x, y, \ldots$ will refer to lattice points, $i, j,
\ldots$, 
will represent flavour indices, while Greek
letters $\alpha, \beta, \ldots$ are used for Dirac indices):

\begin{eqnarray}
\hat M_{x\alpha i;y\beta j}&=&K_{x\alpha i;y\beta j}+Y_{x\alpha
  i;y\beta j},
  \label{matrixM} \\
K_{x\alpha i;y\beta j}&=&\frac{1}{2}\sum_{\mu}\left(\delta_{x+\mu,y}-
\delta_{x-\mu,y}\right)\sigma^{\mu}_{\alpha\beta}\,\delta_{ij},
  \label{MATRIZK} \\
Y_{x\alpha i;y\beta j}&=&y\,\delta_{xy}\left({\mbox{\boldmath $\phi$}}\cdot
{\mbox{\boldmath {$\tau$}}}\right)_{ij}\,\delta_{\alpha\beta}.
  \label{matrixY}
\end{eqnarray}
Keeping in mind that for Pauli matrices $\sigma_2\sigma_i\sigma_2=-\sigma_i^*$,
where $^*$ means complex conjugation, and that 
$\left[{\mbox{\boldmath $\gamma$}},{\mbox{\boldmath $\tau$}}\right]=0$,
one easily proves that, for real $y$,
\begin{equation}
\sigma_2\tau_2\left(K\, +\, Y\right)\sigma_2\tau_2=
-\left(K^*\, +\, Y^*\right).
\label{REALDET}
\end{equation}
Therefore,
\begin{equation}
\det\left(K\, +\, Y\right)=\det\left(-K^*\, -\, Y^*\right)=
\left[\det\left(K\, +\, Y\right)\right]^*,
\end{equation}
{\em i.e.} the determinant is real.
Thus, by doubling the number of fermion families, we obtain a positive
determinant. Had we introduced a chemical potential, $\mu$, the only change
would be the introduction of $\e^{\pm\mu}$ on the temporal links of
the kinetic matrix~\cite{KARSCH}. The essential requirement
for Eq.\ (\ref{REALDET}) to hold (that the only non-real numbers are in 
{\mbox{\boldmath $\gamma$}},{\mbox{\boldmath $\tau$}}) is thus
not endangered by the chemical potential and the determinant is still real. 

\begin{figure}[htb]
\begin{center}
\leavevmode
\centering\epsfig{file=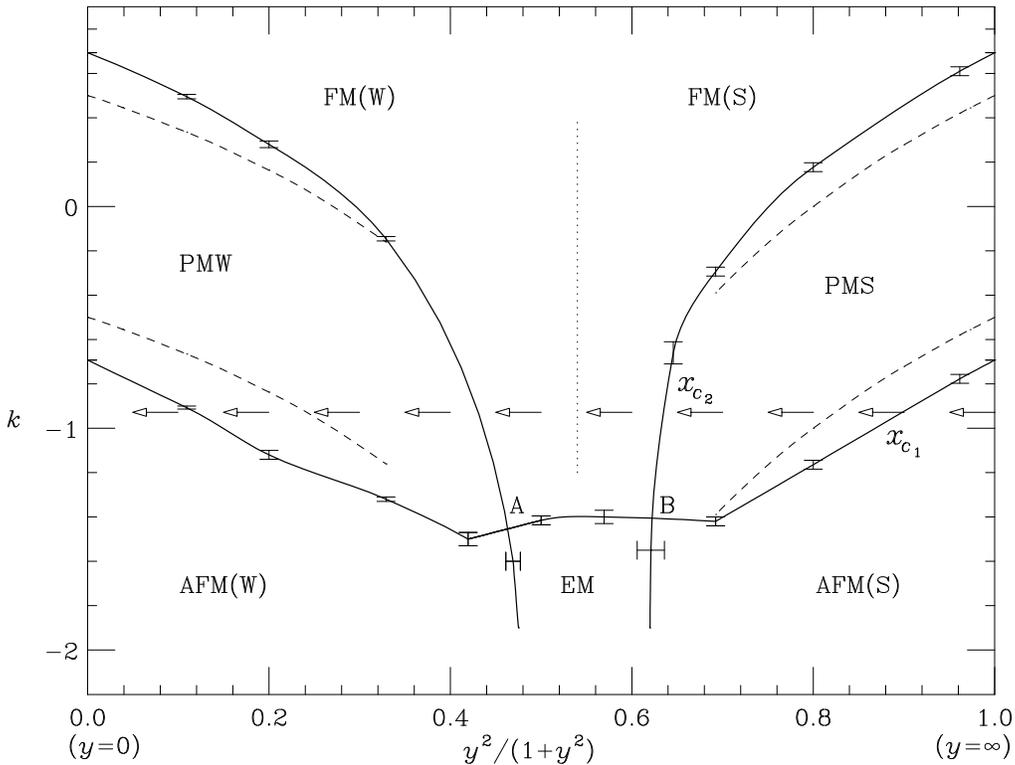,width=0.7\linewidth,angle=90}
\end{center}
\caption{Phase diagram of the action (\protect\ref{ACCION}), for two
fermion families.
Dashed lines are from the MF calculation,
solid lines from a MC calculation on an $8^3$ lattice.
The arrows indicate the doping-``trajectory''.}
\label{PHASES}
\end{figure}

The phase diagram of the model at zero temperature is shown 
in fig.\ \ref{PHASES}. Notice that
it is very
similar to the phase diagram of (chiral) Yukawa models for the electroweak
sector of the Standard Model of elementary particle 
interactions~\cite{shigemitsubocketal}.
At $y=\infty$ and at $y=0$ we
recover the non-linear $\sigma$-model (see sections \ref{MFPD},\ref{MCPD}) with its well known
paramagnetic (PM), ferromagnetic (FM) and antiferromagnetic (AFM) phases. 
At finite $y$, we expect these phases to extend into the ($k$,$y$) plane.
One of its most important features is that there are two mutually
disconnected paramagnetic phases, one at weak coupling (called PMW)
and one at strong coupling (PMS).
One sees that the PMW-FM and the PMW-AFM transition lines meet in a point
{\bf A}, where this disordered phase ends.
In the strong coupling
sector of the phase diagram, a similar behaviour is found, with the
two transition lines meeting at point {\bf B}.
This observation means that one may expect totally different behaviour
in each of the two paramagnetic phases.
This is indeed the case, as we shall see later.

As there is no evidence for a phase transition between the strong- and
weak-coupling regions of the FM and AFM phases,
we name them FM(W) and FM(S), AFM(W) and AFM(S) (note the parentheses).
There may be crossovers between these regions, though.

Between the points {\bf A} and {\bf B}, we find a phase where both
the magnetization and the staggered magnetization are different
from zero. We name this phase {\it exotic magnetic} (EM).
An appealing possibility is that it corresponds to
a helicoidal phase.
This would merit a more detailed study (see section \ref{physzeroT}
for some comments on this point).
We expect the EM phase to disappear for large enough $-k$, but we have not
explored this numerically.

\subsection{Phenomenological use of the model}

\label{PHEN}

Before embarking on a detailed study of the model, let us briefly discuss
its relation with the phenomenology of the cuprates.
In particular, it is important to explain how
the doping fraction $x$ is accounted for by the model.

By increasing the doping fraction, one increases the
overall mobility of the charge carriers.
In our model, this mobility is embodied in the fermionic hopping
parameter $\rho$ in (\ref{ACCIONZ}).
The exact mapping between $x$ and $\rho$ is not important for the qualitative
description, but it is important
that in the limit $x, \rho \rightarrow 0$ one recovers the O(3) model of the
undoped compounds (cf.\ guideline A. in Section \ref{SYM}).
Thus, the undoped compound will be described by $y=\lambda/\rho=\infty$,
and the large-$y$ regime
corresponds to the strongly-correlated small-$x$ region, with immobile,
localized carriers \cite{MURAMATSU2}.

Within the lattice O(3) model, we still have two possible points to describe
the undoped materials (see fig.\  \ref{PHASES}).  One in the FM phase and one 
in the AFM phase,
related by symmetry (\ref{symmZ2}).
Although one usually considers the FM phase, especially in continuum
descriptions, it is perfectly legitimate (see Eqs.\ (\ref{symmZ2},\ref{SIMSTAGG}) and
comments below it) to consider the AFM phase.
In fact, as the doping
is experimentally known to destroy the antiferromagnetism, this turns out
to be the most natural option (see fig.\ \ref{PHASES}).
(Let us remark though, that the reader can start in the ferromagnetic phase
if he/she so wishes, by taking an imaginary coupling constant, as in Eq.\ 
(\ref{symmZ2})).

The most important effect of changing the doping fraction, $x$, will
be a change in the effective spin-fermion coupling, $y$ (indicated by the
arrow-line in figure \ref{PHASES}), and therefore 
a change of
the system ground state and excitations (fermionic and bosonic).
Strictly speaking however, there will be a relation between $x$
and a chemical potential, $\mu$, which in fact could be introduced in
our formulation (recall that the fermion determinant would remain
real). This relation would be in addition to the (phenomenological)
relation between $x$ and $(y,k)$.
We will now argue, though, that for the physics considered in this paper 
we can ignore the chemical potential.
In other words, a possible chemical potential
would not change the excitations of the ground state in any essential way. 

Let us start with a general point, that our model captures the insertion
of charge carriers in an effective way, by increasing the overall mobility.
One could object to this that a vanishing chemical potential entails
the absence of a Fermi surface.
In the strong-coupling region, however, there are two possible responses
to this objection.
First of all,
for small to intermediate $x$, ({\em i.e.} before
and just after superconductivity appears) 
the very concept of a Fermi surface is very problematic from the
experimental point of view~\cite{NORMAN}.
Second, it will turn out that the Fermi surface plays no role in
our mechanism to form bound charge pairs.
Therefore, ignoring a possible chemical potential in the small-$x$ region,
whose main role would have been to fix the Fermi surface, is perfectly
consistent, both with the experimental situation and within the context
of the dynamical pair formation in our model.

The overdoped, large-$x$ region, on the other hand,
is characterized by Fermi liquid behaviour, so
the presence of a chemical potential in the small-$y$ region of
our model will be essential
to determine the thermodynamics of the system.
Since the only statement we want to make concerning this large-$x$
region is the actual emergence of this Fermi liquid behaviour, we can
forget about the chemical potential also in the large-$x$ region,
for our present purposes.  

In the superconducting regime, finally, a chemical potential would 
change to some extent the ground state and the excitations of the system.
However, one expects that in a region of heavy fermions, a quenched region,
this change will be small.  So we can drop the chemical potential here too.

After having discussed the phenomenological use and interpretation of the
model defined by (\ref{ACCIONZ}), we shall proceed to solve it.

\section{Mean Field Calculations of the Phase Diagram}

\label{MFPD}

Our aim in this section is to determine the zero-temperature
phase diagram of the model in the $y$-$k$ plane (cf.\ Fig.\
\ref{PHASES}), using Mean Field
techniques.  These calculations already provide a lot of insight, especially
for the strong coupling region.  They will be contrasted with
numerical simulations for the phase diagram in Sect.\ \ref{MCPD},
and they will be extended to a study of the relevant charged
(quasi-particle) excitations in Sect.\ \ref{sectMFexc}.

Our Mean Field calculations are based on small- and
large-$y$ expansions combined with the saddle point methods described in
Ref.\ \cite{drouffezuber}.
This approach guarantees a systematic expansion in $1/d$, which is
particularly important for operators which are zero to lowest-order.
Our particular method furthermore allows us
to handle (products of) fermionic variables occurring in the expansion
of the fermion determinant in a well-defined way.
These techniques have been developed and applied in the context of similar
lattice models \cite{plzar,class} of the Electroweak sector of the Standard
Model of elementary particle interactions.

We shall first concentrate on the small-$y$ region, and
incorporate the fermion determinant up to ${\cal O}(y^2)$.

In order to apply the saddle-point method, the integration over the
fields must be unrestricted.
We therefore need to replace the integration over the spin
vectors ${\mbox{\boldmath $\phi$}}$, constrained by the condition $|{\mbox{\boldmath $\phi$}}| = 1$, with an
integration over unconstrained variables ${\mbox{\boldmath $\xi$}}$.
This is done by multiplying the functional integrand in Eq.\ (\ref{Z}) by
\begin{eqnarray*}
1 &=& \int\! D\xi \, \delta({\mbox{\boldmath $\phi$}} - {\mbox{\boldmath $\xi$}})
\equiv \prod_n \prod_{a=1}^3 \int_{-\infty}^\infty \! d\xi_x^a
\, \delta(\phi_x^a - \xi_x^a)
\\ 
 &=&
\prod_x \prod_{a=1}^3 \int_{-\infty}^\infty \! d\xi_x^a
 \, \int_{-\infty}^\infty \! \frac{dA_x^a}{2\pi}
 \, \exp[iA_x^a (\phi_x^a - \xi_x^a)]\ ,
\end{eqnarray*}
and replacing a conveniently chosen subset of the ${\mbox{\boldmath $\phi$}}$ variables
in the action $S$ with ${\mbox{\boldmath $\xi$}}$ fields.
We obtain
\begin{eqnarray}
\label{Z2}
Z &=&
\int \! \frac{D\xi DA}{(2\pi)^3}
 \, \exp\left[ k \sum_{x,\mu} {\mbox{\boldmath $\xi$}}_x \cdot {\mbox{\boldmath $\xi$}}_{x+\mu}
 - i \sum_x {\mbox{\boldmath $A$}}_x \cdot {\mbox{\boldmath $\xi$}}_x \right]
 \nonumber \\
 &\times &
\int \! D\bar\psi D\psi
 \, \exp\left[ -\sum_{x,\mu} \frac12 \bar\psi_x \gamma^\mu (\psi_{x+\hat\mu} -
    \psi_{x-\hat\mu}) \right]\\\nonumber
 &\times & \prod_x \left\{ \int \! \frac{d{\mbox{\boldmath $\phi$}}_x}{4\pi}
 \, \exp\left[ i{\mbox{\boldmath $A$}}_x \cdot {\mbox{\boldmath $\phi$}}_x
        - y \, \bar\psi_x \, {\mbox{\boldmath $\phi$}}_x\cdot\vec\tau \, \psi_x \right] \right\}
 . 
\end{eqnarray}
Note that both the ${\mbox{\boldmath $\xi$}}$ fields and the auxiliary fields ${\mbox{\boldmath $A$}}$ are
unconstrained.

Now we have to integrate out the constrained variables
$\phi^a_n$ (as well as the fermions), before the mean fields can be introduced.
Let us concentrate on a single ${\mbox{\boldmath $\phi$}}_n$ integration in Eq.\ (\ref{Z2}),
dropping the subscripts $n$ for simplicity.
First, we perform an expansion in powers of $y$.
We can write
\begin{eqnarray}
&&\int \! \frac{d{\mbox{\boldmath $\phi$}}}{4\pi}
 \, \exp\left[ i{\mbox{\boldmath $A$}} \cdot {\mbox{\boldmath $\phi$}}
        - y \, \bar\psi \, {\mbox{\boldmath $\phi$}}\cdot\vec\tau \, \psi \right]
 \nonumber \\
&&=
\exp\left[ u(i{\mbox{\boldmath $A$}}) \right]
 \, \exp\left[ -y\, Q^a \cdot \langle\phi^a\rangle_{i{\mbox{\boldmath $A$}}}
  + \frac12 y^2 \, Q^a Q^b T^{ab} + {\cal O}(y^3) \right]
  , \label{zphi2}
\end{eqnarray}
where we have defined
$$Q^a = \bar\psi \tau^a \psi\ \  ,\ \  
u(i{\mbox{\boldmath $A$}}) = \ln \, \int \! \frac{d{\mbox{\boldmath $\phi$}}}{4\pi}
\, \exp\left[ i{\mbox{\boldmath $A$}} \cdot {\mbox{\boldmath $\phi$}} \right]
\ ,\  T^{ab} =\langle \phi^a \phi^b\rangle_{i{\mbox{\boldmath $A$}}} - 
\langle\phi^a\rangle_{i{\mbox{\boldmath $A$}}}
\langle\phi^b\rangle_{i{\mbox{\boldmath $A$}}},$$
and we have introduced the notation
$$
\langle{O}\rangle_{i{\mbox{\boldmath $A$}}} =
\left.{\int \! \frac{d{\mbox{\boldmath $\phi$}}}{4\pi}}
\, O \, \exp\left[ i{\mbox{\boldmath $A$}} \cdot {\mbox{\boldmath $\phi$}} \right] \right/
{\int \! \frac{d{\mbox{\boldmath $\phi$}}}{4\pi}
\, \exp\left[ i{\mbox{\boldmath $A$}} \cdot {\mbox{\boldmath $\phi$}} \right] }
\, .$$

In addition, we introduce a Hubbard-Stratonovich vector parameter
${\mbox{\boldmath $\lambda$}}$ to deal with the quartic fermion term in Eq.\ (\ref{zphi2}),
\begin{eqnarray}
\exp\left[ \frac12 y^2 \, \sum_{a,b} \, Q^a Q^b T^{ab}\right]
&=&
\int \! \frac{d{\mbox{\boldmath $\lambda$}}}{(2\pi)^{3/2}}
 \, \exp\left[ - \frac12 \sum_a \lambda^a \lambda^a
\right]\\\nonumber
&\times&\exp\left[y \sum_{ab} \left(\sqrt{T}\, \right)^{ab}
    Q^a \lambda^b\right]\, . 
\label{HSlambda}
\end{eqnarray}
(Note that the matrix  $T$ is self adjoint, and positive definite if 
${\mbox{\boldmath $A$}}$ is imaginary, so the square root is well defined). 
Thus, up to this order in $y^2$,
the action is bilinear in the fermion fields.

Carrying out the fermion integration in Eq.\ (\ref{Z2}) now gives det $M$,
where
\begin{equation}
M_{x,\alpha,i;y,\beta,j} = K_{x,\alpha,i;y,\beta,j}
  + y \delta_{xy} \delta_{\alpha\beta} \sum_a
 (\langle \phi^a_x \rangle_{iA_x} - 
\sum_b\left(\sqrt{T_x}\right)^{ab} \lambda^b) 
\tau^a_{ij}
 \, . \label{Mdef}
\end{equation}
The matrix $K$ has been defined in Eq. (\ref{MATRIZK}).

The mean fields are the field values at the saddle point of the free energy
\begin{equation}
-F = \sum_x u(i{\mbox{\boldmath $A$}}_x) + k \sum_{x,\mu} {\mbox{\boldmath $\xi$}}_x \cdot {\mbox{\boldmath $\xi$}}_{x+\mu}
 - i \sum_x {\mbox{\boldmath $A$}}_x \cdot {\mbox{\boldmath $\xi$}}_x
 - \frac12 \sum_x {\mbox{\boldmath $\lambda$}}_x^2 + \mathrm{Tr\,} \log M
 \, . \label{Fdef}
\end{equation}
A choice of the mean fields should be done at this point, as we cannot
calculate $\log M$ for general $\{{\mbox{\boldmath $A$}}_x\, ,\, 
{\mbox{\boldmath $\lambda$}}_x\}$. An appropriate choice for the study
of a PM-FM phase transition is

\begin{eqnarray}
{\mbox{\boldmath $A$}}_x &=& (0,0,-i\alpha)
 \, , \\\nonumber
{\mbox{\boldmath $\xi$}}_x &=& (0,0,v)
 \, ,  \\\nonumber
{\mbox{\boldmath $\lambda$}}_x &=& (0,0,\lambda)
 \, , 
\end{eqnarray}
in terms of which ($N$ is the lattice volume)
\begin{equation}
F/N = - u(\alpha) - kdv^2 + \alpha v + \frac12 \lambda^2
    - \frac1N\, \mathrm{Tr\,} \log M
 \, , \label{FoverN}
\end{equation}
with $\alpha$, $v$ and $\lambda$ satisfying
the saddle point equations
\begin{equation}
\left.\nabla F\, \right|_{(\alpha,v,\lambda)}=0
 \, . \label{sadd}
\end{equation}

The fermion matrix, $M(\alpha,v,\lambda)$, can be calculated
in momentum space, where it is diagonal in its momentum indices.
One easily finds 
\begin{equation}
\mathrm{det\,} M = \exp\left[ 2 \sum_p \log \frac{ \sum_{\mu=1}^3 \sin^2 p_\mu
  + y^2 \left(u'(\alpha) - \lambda \sqrt{u''(\alpha)}\right)^2 }
    {\sum_{\mu=1}^3 \sin^2 p_\mu }  \right]
 \, , \label{Mresult}
\end{equation}
where we have divided out the determinant for free fermions.
We need only the leading ${\cal O}(y^2)$ contribution to the exponent,
hence the mean field free energy becomes, in the infinite volume limit:
\begin{equation}
F/N = - u(\alpha) - kdv^2 + \alpha v + \frac12 \lambda^2
    - 2 y^2 \left(u'(\alpha) - \lambda \sqrt{u''(\alpha)}\right)^2 \, C_0
 \, , \label{FoverN2}
\end{equation}
where
\begin{equation}
C_0 \ =\ \int_{-\pi}^\pi \! \frac{d^3p}{(2\pi)^3}
 \, \frac1{\sum_{\mu=1}^3 \sin^2 p_\mu} \ =\ 1.0109240
 \, . \label{C0}
\end{equation}

Next, we shall discuss the actual solutions to Eqs.\ (\ref{sadd}).
From
$u(\alpha) \ =\ \ln \, (\sinh\alpha / \alpha)$, one easily 
finds that $\alpha = v = \lambda = 0$ always fulfill them.
For small $k$, $y$, it is a true minimum
of the free energy.
This characterizes a paramagnetic (PM) phase, since none of the fields
develops an expectation value.

For larger values of $k$ and $y$, there is another, non-trivial solution,
corresponding to a ferromagnetic (FM) phase.
It emerges when a negative mode in $F/N$ starts to develop, as a function
of the mean fields, and the
transition between the two regions is given by the condition ($F''$ is
the Hessian matrix)
\begin{equation}
\left.\mathrm{det\,}F''\right|_{(\alpha=0,v=0,\lambda=0)} = 0
 \, . \label{transcond}
\end{equation}
This condition is satisfied for $F/N$ of Eq.\ (\ref{FoverN2})
if 
\begin{equation}
k = \frac3{2d} - \frac{2\, C_0}d y^2
 \, . \label{ky2}
\end{equation}

This curve defines the phase transition line between the PM and FM
phases in the small-$y$ region.
Using the symmetry (\ref{symmZ2}), we deduce that there is a similar
transition separating the PM and AM phases,
\begin{equation}
k = - \frac3{2d} - \frac{2\,C_0}d y^2
 \, . \label{ky2a}
\end{equation}
Let us finally remark that in the presence of $N_f$ such fermion fields
we would have $N_f$ factors of det~$M$, leading to a multiplication of
$C_0$ by $N_f$ in Eqs.\ (\ref{ky2}) and (\ref{ky2a}).

The large-$y$ region is easier to deal with.
Here it is convenient to integrate out the fermions directly
in Eq.\ (\ref{Z2}), leading to (summation over repeated index is carried)
\begin{eqnarray}
\mathrm{det\,} M_{x,\alpha,i;y,\beta,j} &=&
\mathrm{det\,} \left(K_{x,\alpha,i;y,\beta,j}
+ y \delta_{\alpha\beta} \delta_{xy} \sum_a \phi^a_x \tau^a_{ij} \right)
 \label{Mylarge0} \\
&=&
\mathrm{det\,} \left(y\delta_{\alpha\gamma} \delta_{xz} \sum_a \phi^a_x
  \tau^a_{ik} \right) \nonumber \\
&&\times\
\mathrm{det\,} \left( \delta_{zy} \delta_{\gamma\beta} \delta_{kj}
 + \frac1y \sum_b \phi^b_x \tau^b_{kl} K_{z,\gamma,l;y,\beta,j}  \right)
 \, . \label{Mylarge}
\end{eqnarray}
Here we have used that $(\sum_a \phi^a \tau^a)^2 = \openone$ (recall
the ${\mbox{\boldmath $\phi$}}$'s are unit vectors).
Now we can expand log(det $M$) in powers of $1/y$.
The ${\cal O}(1/y)$ term vanishes by virtue of $K_{xx} = 0$.
To second order one obtains
\begin{eqnarray}
\log \mathrm{det\,} M &=& \log y^{4N} + \mathrm{Tr\,} \left(
 -\frac1{2y^2} \sum_a \phi^a_x \tau^a_{ki} K_{x\alpha i;t\gamma l}
\sum_b \phi^b_t \tau^b_{lj} K_{t\gamma j;y\beta p}\right)
  \label{trlogM0} \\
&=&
\log y^{4N} + \frac1{y^2} \sum_{x,\mu}
    {\mbox{\boldmath $\phi$}}_x \cdot {\mbox{\boldmath $\phi$}}_{x+\hat\mu}
 \, . \label{trlogM2}
\end{eqnarray}
Here, $\log y^{4N}$ is an irrelevant constant that can be
dropped. Notice also that
this expression will acquire a prefactor $N_f$ if there are $N_f$ identical
fermion flavours.
One sees that, up to ${\cal O}(1/y^2)$, the only effect of the fermion
determinant is a renormalization of the scalar hopping parameter of the
O(3) model,
\begin{equation}
k \rightarrow k + N_f \, \frac1{y^2}
  \, . \label{kRenorm}
\end{equation}

Note that we did not introduce any mean fields to derive this result.
The usual MF treatment of the O(3) model with this renormalized coupling
now immediately gives us the required phase transition lines in the large-$y$
region of our model:
\begin{equation}
k = \pm \frac3{2d} - N_f \, \frac1{y^2}
  \, . \label{ky2b}
\end{equation}

It is interesting to compare the small- and large-$y$ results,
to leading order in $1/d$.
As is well known, the first order in this expansion is equivalent to
any MF approximation, up to
higher-order terms.
For this purpose, we need the $1/d$ expansion of the constant $C_0$
in Eq.\ (\ref{C0}), which can be calculated as follows:
\begin{eqnarray}
C_0(d) &=& \int_{-\pi}^\pi \! \frac{d^dp}{(2\pi)^d}
 \, \frac1{\sum_{\mu=1}^d \sin^2 p_\mu} 
\ =\ 
2 \int_0^\infty \! ds \, (e^{-s} I_0(s))^d
 \label{C0exp2} \\
&=& \frac2d \, \left(1 + \frac1{2d} + {\cal O}\left(\frac1{d^2}\right) \right)
 \, , \label{C0exp3}
\end{eqnarray}
where $I_0(s) = \int_{-\pi}^\pi (d\theta/2\pi) \, \exp(s\cos\theta)$
is the modified Bessel function.
In fact, the second equality in  Eq.\ (\ref{C0exp2}) was used to obtain
the numerical result (\ref{C0}) for $C_0$.

Keeping only the leading-order term $2/d$ for $C_0$ we find that the
phase transition lines would meet at $y^2 =2/d$.

Now we are ready to map out the phase diagram of the model, as predicted
by the MF method for the weak and strong coupling regions.
This is done in Fig.\ \ref{PHASES}.
The vertical axes at $y=0$ and $y=\infty$ correspond to the O(3) model,
with its disordered (PM) and ordered (FM and AFM) phases.
These phases extend into the $y$-direction, both for $y>0$ and $y<\infty$.
Note that all the phase transition lines bend downward.
This can be understood intuitively by assuming a MF value for the
fermion condensate, which would act as an external field tending to
align the spins $\phi$ in parallel.

\section{Monte Carlo: Method and Results}

\label{MCPD}

A well established method for dynamical fermion simulations is Hybrid
Monte Carlo (HMC)~\cite{HMC}. However, the implementation of this
algorithm in a model with constrained variables is not
straightforward.  This has been satisfactorily achieved for models
with variables belonging to a Lie group~\cite{LIE}, like SU($N$) gauge
theories or like some spin-models, such as the O($N=2,4$) non-linear
$\sigma$-models. However, for other spin variables (not in a Lie group), as in the 
O(3) non-linear $\sigma$-model, this had not been satisfactorily  
solved yet, although the problem arose already in the first
simulations using the Langevin algorithm \cite{PARISI}.
Our solution is a generalization
of the strategy in~\cite{LIE}.

We shall first
discuss our solution in the quenched approximation, where comparison
with other algorithms is possible (Section \ref{QUENCHED}), 
and then deal with the full theory in Section \ref{FULL}. 
Finally our Monte Carlo results for the phase diagram of the full
theory will be presented in Section \ref{NUMERICAL}.

\subsection{The HMC method for the quenched approximation}

\label{QUENCHED}

For the purpose of discussion it will prove convenient to briefly
describe the HMC method for unconstrained bosonic variables $\phi(x)$ with
action $S_B(\phi)$ (see ref.~\cite{HMCBOOK} for a pedagogical
presentation):

\begin{enumerate}
\item Introduce uncorrelated gaussian variables $\pi(x)$ of unit variance
(the {\it conjugate momenta} for the fields $\phi$) and
define a Hamiltonian 
\begin{equation}
H=\sum_x \frac{1}{2}\pi^2(x)+S_B(\phi)
 \, . \label{Ham}
\end{equation}
One can then use the hamiltonian equations of motion
\begin{eqnarray}
{\dot\phi}(x,\tau) &=& \pi(x,\tau) \, , \label{eom1} \label{eom2}\\
{\dot\pi}(x,\tau) &=& -\frac{\delta S_B}{\delta \phi(x,\tau)} \, , \nonumber
\label{TEQ}
\end{eqnarray}
to perform a microcanonical Molecular Dynamics evolution in
``Monte Carlo time'', $\tau$.
After a certain period of MC time (called ``trajectory''),
new random momenta $\pi(x)$ are chosen (``refreshing'' the momenta).
The crucial properties of  Eqs.\ (\ref{TEQ}) are their 
time reversibility, 
and the invariance of the Liouville measure, 
$D \phi\, D \pi$, under time evolution.

\item
In practice, the molecular dynamics equations of motion for a trajectory are
discretized into $N$ steps $\Delta\tau$.  This is done using a
leap-frog algorithm which is {\it exactly} time reversible, but does
introduce a systematic error which shows up as a non-zero $\Delta H =
{\cal O}(\Delta\tau^2)$.  The {\it detailed-balance} is not endangered
by this error, because a Metropolis acceptance step is performed.  For
fixed trajectory length, $N$ can then be tuned to optimize the overall
efficiency.

\end{enumerate}

To generalize the method to constrained variables, one needs to
appropriately define the conjugate momenta and the equations of motion
in order to preserve the constraint and, most importantly, not to
spoil the time reversibility. Each spin variable,
${\mbox{\boldmath$\phi$}}$, lives on the surface of a two-sphere, and
correspondingly one could imagine an algorithm with two independent
conjugate momenta, maybe living in the perpendicular plane
(${\mbox{\boldmath$\phi$}}\cdot{\mbox{\boldmath$\pi$}}=0$). However,
changing the constraint from the field ${\mbox{\boldmath$\phi$}}$ to
the momenta is not very appealing (and, from the practical side, one
would need to worry about {\it two} constraints in the numerical
integration).
A different approach, the use of
spherical coordinates, has the drawback of a non-planar integration
measure. Our very simple algorithm avoids constraints and non-planar
measures, by introducing {\it three} conjugate momenta per spin.

We shall start from an analogy with the
dynamics of a particle living in the sphere, a potential ($V$) acting on it.
The Hamiltonian is
\begin{equation}
H^{\mathrm{sphere}}=\frac{{\mbox{\boldmath$L$}}^2}{2} + V({\mbox{\boldmath$\phi$}}).
\label{HSPHERE}
\end{equation}
Here ${\mbox{\boldmath$L$}}$ is the angular momentum,
${\mbox{\boldmath$\phi$}}\times{\dot{\mbox{\boldmath$\phi$}}}\,$.
The equations of motion can now be obtained from the Poisson
Bracket~\cite{GOLDSTEIN} with the hamiltonian (\ref{HSPHERE}):
\begin{equation}
{\dot{{\mbox{\boldmath $\phi$}}}}\ =\ 
{{\mbox{\boldmath $L$}}}\times{{\mbox{\boldmath $\phi$}}}\ ,\ 
{\dot{{\mbox{\boldmath $L$}}}}\ =\ -{{\mbox{\boldmath $\phi$}}}\times
\frac{\delta V}{\delta {{\mbox{\boldmath $\phi$}}}}.
\label{EQSPHERE}
\end{equation} 
In this expression $\frac{\delta V}{\delta{{\mbox{\boldmath $\phi$}}}}$
stands for $\left(\frac{\delta V}{\delta\phi_1},
\frac{\delta V}{\delta\phi_2},\frac{\delta V}{\delta\phi_3}\right)$.
 
This formalism is still
inconvenient for us, because the constraint
${\mbox{\boldmath$\phi$}}\cdot{\mbox{\boldmath$L$}}=0$ complicates the
generation of random momenta according
to a Gaussian distribution.
However, the following simple facts can be straightforwardly
established from the equations (\ref{EQSPHERE}):

{\bf I.}~Both ${\mbox{\boldmath$\phi$}}^2$ and 
${\mbox{\boldmath$\phi$}}\cdot{\mbox{\boldmath$L$}}$ are conserved
through the time evolution. If the initial condition verifies the
constraints
${\mbox{\boldmath$\phi$}}^2=1\ ,\ 
{\mbox{\boldmath$\phi$}}\cdot{\mbox{\boldmath$L$}}=0$, this
will not be spoiled by the dynamics.

{\bf II.}~The dynamics is time-reversible.

{\bf III.}~Although the $L_i$ cannot be
all canonical variables~\cite{GOLDSTEIN},
the ``Liouville'' measure, 
$D {\mbox{\boldmath$\phi$}}\, D {\mbox{\boldmath$L$}}(=
d \phi_1\, d \phi_2\, d \phi_3\ d L_1\, d L_2\, d L_3)$,
is left invariant by the time-evolution.

{\bf IV.}~The Hamiltonian is a constant of the motion.\\ Now let us forget
about the constraint
${\mbox{\boldmath$\phi$}}\cdot{\mbox{\boldmath$L$}}=0$, {\em i.e.} we
introduce a new field ${\mbox{\boldmath$P$}}$ which can have a
``radial component'' (it is no longer an angular momentum), but we keep
the Eqs. of motion (\ref{EQSPHERE}). Obviously, statements {\bf
I}--{\bf IV} will still hold. Whether a symplectic structure
is hidden under this new dynamical system is unclear, but also
irrelevant (properties {\bf II} and {\bf III} are the essential ones
for HMC to be a correct algorithm~\cite{HMCBOOK}).

So, we introduce three momenta per spin, ${\mbox{\boldmath $P$}}=(P_1,P_2,P_3)$,
and write down the Hamiltonian
\begin{equation}
H \ =\ \sum_{x}\frac{{\mbox{\boldmath $P$}}^2}{2} +
       S_B({\mbox{\boldmath $\phi$}}).
\label{HQUENCHED}
\end{equation}
Equations of motion  respecting properties {\bf I}--{\bf IV} are easily generalized:
\begin{equation}
{\dot{{\mbox{\boldmath $\phi$}}}}_{(x,\tau)}  \ =\ 
   {{\mbox{\boldmath $P$}}}_{(x,\tau)}
   \times{{\mbox{\boldmath $\phi$}}}_{(x,\tau)}\  ,\ 
   {\dot{{\mbox{\boldmath $P$}}}}_{(x,\tau)}  \ =\ 
   -{{\mbox{\boldmath $\phi$}}}_{(x,\tau)}\times
   \frac{\delta S_B}{\delta{{\mbox{\boldmath $\phi$}}}_{(x,\tau)}} \, .
\label{EQQUENCHED}
\end{equation}
As expected, the evolution equations for the $S^2$ fields $\phi$
take the form of (infinitesimal) rotations, while the conjugate momenta 
can be considered as living in the Lie Algebra of SO(3).
The discretized leap-frog form of these equations is therefore naturally
formulated in terms of finite SO(3) rotations,
\begin{eqnarray}
{\mbox{\boldmath $\phi$}}_x(n\Delta\tau+\Delta\tau)&=&
\exp[\Delta\tau{\mbox{\boldmath $P$}}_x((n+\frac{1}{2})\Delta\tau)\cdot
{\mbox{\boldmath $J$}}] \,
{\mbox{\boldmath $\phi$}}_x(n\Delta\tau)
   \, , \label{eom1a}\\
{\mbox{\boldmath $P$}}_x((n+\frac{1}{2})\Delta\tau)&=&
{\mbox{\boldmath $P$}}_x((n-\frac{1}{2})\Delta\tau)\, -\,
{{\mbox{\boldmath $\phi$}}}_{(x,n\Delta\tau)}\times
\frac{\delta S_B}{\delta{{\mbox{\boldmath $\phi$}}}_{(x,n\Delta\tau)}}
     \Delta\tau
   \, , \label{eom2a}
\end{eqnarray}
where ${\mbox{\boldmath $J$}}$ are the $3\,\times\,3$ generators of SO(3),
satisfying
\begin{equation}
\left(\exp[\theta{\mbox{\boldmath $n$}}\cdot{\mbox{\boldmath $J$}}]\right)_{ij}
 \ =\ 
  \delta_{ij}\,\cos\theta+n_in_j\,(1-\cos\theta)-\epsilon_{ijk}n_k\,\sin\theta
 \label{Jchoice}
\end{equation}
for unit vectors ${\mbox{\boldmath $n$}}$.
Again, the length constraint on the $\phi$ fields is preserved by construction.

This final result is reminiscent of the elegant solution
for models with variables belonging to a Lie group and conjugate
momenta in the group algebra (or vice versa) \cite{LIE}.

\begin{table}
\caption{Values for several observables in the quenched model (\ref{ACCION})
on an $8^3$ lattice at $k=0.693 \approx k_c$, obtained with our implementation
of HMC and with Wolff's single cluster algorithm \protect\cite{WOLFF}.}
\medskip
\hrule\hrule
\begin{tabular*}{\hsize}{@{\extracolsep{\fill}}llllll}
Algorithm & \multicolumn{1}{c}{$\langle E \rangle$}
& \multicolumn{1}{c}{$\partial_k\langle E \rangle$}
&\multicolumn{1}{c}{$\chi/V$}&\multicolumn{1}{c}{$\xi$}
&\multicolumn{1}{c}{$B$}\\
\hline
HMC   & 0.3505(5) &  1.51(2)  & 0.1426(9)   & 4.47(2) & 0.800(6) \\
Wolff & 0.35061(13) & 1.501(10) & 0.1432(2) & 4.486(9) & 0.8031(18)
\end{tabular*}
\hrule\hrule
\label{TABLAQ}
\end{table}

In our case, $S_{B\, \mathrm{quenched}}=-k \sum_{n,\mu} \, {\mbox{\boldmath
$\phi$}}_n\cdot {\mbox{\boldmath $\phi$}} _{n+\hat\mu}$, so the HMC
algorithm can now be implemented in a straightforward manner.  To test the
algorithm,
we have simulated the O(3) model on an $8^3$ lattice at
$k=0.693 \approx k_c$ \cite{O3} with our
HMC algorithm and with Wolff's single-cluster embedding algorithm
\cite{WOLFF}. Let us first define the measured observables,
and then compare them.

In this work we have only measured bosonic observables, as our sole objective
was the numerical determination of the phase diagram. We have 
constructed our observables in terms of the Fourier transform
of the spin field:
\begin{equation}
\widehat{m}(\mbox{\scriptsize \boldmath $p$}) \ =\
   \frac{1}{V}\sum_{\mbox{\scriptsize \boldmath $x$}}
    \exp(-i{\mbox{\boldmath $p$}}\mathbf{\cdot} {\mbox{\boldmath $x$}})\ 
    {{\mbox{\boldmath $\phi$}}}_{\mbox{\scriptsize \boldmath $x$}} \, ,
  \label{mdef}
\end{equation}
where $V = L^3$ is the lattice volume.

We define the non-connected finite-volume susceptibilities as
\begin{equation}
{\mbox{$\chi$}}=
V \left\langle \widehat{m}^2 (0,0,0)\right\rangle , \quad
\quad {\mbox{$\chi$}}_\mathrm{s}= 
V \left\langle \widehat{m}^2 (\pi,\pi,\pi)\right\rangle.
  \label{susc}
\end{equation}

The subscript `s' on $\chi_\mathrm{s}$ stands for `staggered',
and this term is used to label quantities which are taken with a weight
$-1$ for the odd lattice sites, corresponding to momentum $(\pi,\pi,\pi)$.
Notice that
${\mbox{$\chi$}}/V$ is a pseudo order parameter, which should
be of order one in a ferromagnetically broken phase, and of
order $1/V$ in a paramagnetic or antiferromagnetic phase (and similarly for
${\mbox{$\chi$}}_\mathrm{s}/V$).

Another quantity of interest is the Binder cumulant

\begin{equation}
B=\frac{5}{2}-\frac{3}{2}
\frac{\left\langle\left( \widehat{m}^2 (0,0,0)\right)^2\right\rangle}
{\left\langle \widehat{m}^2 (0,0,0)\right\rangle^2} \, ,
  \label{Bdef}
\end{equation}
with an analogous definition for the staggered variant $B_\mathrm{s}$.

One expects $B=1$ in the FM phase, where
${\mbox{$\chi$}}/V$ is non-vanishing in the thermodynamic limit, while
it should be of order $1/V$ in the PM phase, far from the phase transition.

For the correlation length, we use a definition which 
is easy to measure and gives accurate results: 
\begin{equation}
\xi=\left(\frac{\chi/F-1}{4\sin^2(\pi/L)}\right)^{1/2},
\label{XI}
\end{equation}
where  $F$ is the squared Fourier transform at minimal 
non-zero momentum,
\begin{equation}
F \ =\ \frac{V}{3}\left(\left\langle\left | \widehat{m}
      (2\pi/L,0,0)\right |^2\right\rangle
      \ + \  \mathrm{permutations}\right) \, .
\label{secondFdef}
\end{equation}
Again, the generalization to staggered quantities is straightforward.
Another kind of observable, needed for the standard
extrapolation method~\cite{FS}, is the normalized nearest-neighbour energy
\begin{equation}
E\ =\ \frac{1}{3V}\sum_{x,\mu} \, \left\langle{\mbox{\boldmath
  $\phi$}}_x\cdot {\mbox{\boldmath $\phi$}}_{x+\hat\mu}\right\rangle
 \ =\ \frac\partial{\partial k} \ln Z \, .  \label{Enn}
\end{equation}

We also measure its fluctuation, given by 
\begin{equation}
3V \left( \left\langle E^2 \right\rangle - \left\langle E \right\rangle^2\right)
\ = \ \frac\partial{\partial k}\left\langle E \right\rangle 
 \, . \label{Ennfluct}
\end{equation}

In Table~\ref{TABLAQ} we compare the values obtained for these observables,
using our HMC algorithm and the single-cluster algorithm.
We find excellent agreement.
Of course the efficiency of our implementation of HMC is not competitive with
a cluster method in the O(3) non-linear $\sigma$-model.
But it could be useful in other
models where cluster methods are not effective in reducing the
dynamical critical exponent $z$ (for instance, when some kind of 
frustration is present \cite{NOCLUSTER}), while HMC
is expected to yield $z=1$ for any bosonic model.

\subsection{The HMC algorithm for the full theory}

\label{FULL}

The only restriction imposed on HMC is that the fermion bilinear
in the action should be given in terms of a positive definite matrix.
This will the case if we consider two identical fermion families ($N_f=2$)
as is usually done in lattice gauge theories.
After integrating them out we obtain $(\det \hat M)^2 =
\det ({\hat M}^{\dag} \hat M$),
where $\hat M$ is the fermion matrix for a single fermion family.
As we are mainly interested in the strong spin-fermion coupling region, 
it makes sense to perform the following manipulation:
\begin{equation}
\det\, \hat M \ =\ \det\, (Y+K) \ =\ y^{4V}\,\det\, (1+Y^{-1}\!K)
\label{STRONG}
\end{equation}
(cf.\ Eqs.\ (\ref{Mylarge0},\ref{Mylarge})).
The constant factor $y^{4V}$ can be dropped, and we define $M=1+Y^{-1}\!K$.

Next, one re-exponentiates the (inverse) fermion matrix
by introducing the so-called {\em pseudo-fermions\/} $z_x$,
which are complex four-component c-number fields.
The partition function is then
\begin{equation}
Z \ =\ \int\!D{\mbox{\boldmath $\phi$}} \, D\bar z \, Dz \, 
  \exp\left(-S_B-\bar z ({M}^{\dag}  M)^{-1} z\right).
\end{equation}
For further details we refer to Ref.\ \cite{HMCBOOK}.

Now the HMC Hamiltonian becomes
\begin{equation}
H=\sum_x \frac{1}{2}{\mbox{\boldmath $P$}}^2_x\ -\ k\,
\sum_{x,\mu}{\mbox{\boldmath $\phi$}}_x\cdot{\mbox{\boldmath $\phi$}}_{x+\mu}
\ +\ z^{\dag}\left(M^{\dag} M\right)^{-1} z \, ,
\label{FULLH}
\end{equation} 
and the time reversible, constraint and energy preserving equations
of motion are
\begin{eqnarray}
\label{FULLEQ}
{\dot{{\mbox{\boldmath $\phi$}}}}_{(x,\tau)}&=&{{\mbox{\boldmath $P$}}}_{(x,\tau)}\times{{\mbox{\boldmath $\phi$}}}_{(x,\tau)},\\\nonumber
{\dot{{\mbox{\boldmath $P$}}}}_{(x,\tau)}&=&
-k\sum_\mu \left({\mbox{\boldmath $\phi$}}_{(x+\mu,\tau)}+
{\mbox{\boldmath $\phi$}}_{(x-\mu,\tau)}\right)\times
{\mbox{\boldmath $\phi$}}_{(x,\tau)}\\\nonumber  & &
-z^{\dag}\left(M^{\dag} M\right)^{-1}\left[\left(
\frac{\delta M^{\dag}}{\delta{{\mbox{\boldmath $\phi$}}}_{(x,\tau)}}\times
{\mbox{\boldmath $\phi$}}_{(x,\tau)}\right)M\ +\ h.c.\ \right]
\left(M^{\dag} M\right)^{-1}z.
\end{eqnarray}

For the inversion of the fermionic matrix, we have employed 
the conjugate gradient algorithm. To formulate the stopping criterium,
let us define $h = \left(M^{\dag} M\right)^{-1}z$, $h_n$ being the 
$n^{\mathrm {th}}$ trial solution.
We continued the conjugate gradient iteration until
\begin{equation}
\frac{\left| \left(M^{\dag} M\right)h_n\ -\ z \right|^2}{|h_n|^2}\leq R.
\label{STOP}
\end{equation} 

In the simulation, we need the inverse matrix both for the leap-frog
and for the Metropolis accept-reject step. It is clear that $R$ does
not need to be the same in both cases. For the Metropolis step, lack of
accuracy in the inversion will bias the simulation. To control this,
we have checked that the Creutz parameter $\left\langle\exp(-\Delta
H)\right\rangle$ equals $1$ within errors.  In some regions of
parameter space $R$ values as small as $10^{-25}$ were needed.  The
essential requirement on the leap-frog is full reversibility in the
numerical integration of the equations of motion (up to the numerical
precision reachable with 64-bit floating point arithmetic). As first
noticed in ref.\ \cite{GUPTA}, this has no relation with $R$ if the
seed for the conjugate-gradient algorithm is chosen to depend on
the {\it actual} configuration only ($h_0=z$, for instance). However, if
$R$ is too large, the numerical integration will produce large changes
in the Hamiltonian, and the Metropolis acceptance will be poor. We
have found that $R=10^{-7}$ during the leap-frog
steps allows for a  $50\%$ acceptance.

In a first implementation of a new MC algorithm, some consistency
checks are extremely useful.  In addition, there are
three parameters to be adjusted for optimal performance, $N$,$\tau$ and
$R$. We have carried out the following tests:
\begin{enumerate} 
\item We have explicitly checked reversibility of the leap-frog algorithm.
\item We have checked that $\left\langle\exp(-\Delta H)\right\rangle = 1$
within errors.
\item The gaussian expectation values,
$\left\langle z^{\dag}\left(M^{\dag} M\right)^{-1}z\right\rangle=4$ and 
$\left\langle{\mbox{\boldmath $P$}}^2\right\rangle=3$ have been checked.
\item We have checked that $\Delta H\propto (\Delta\tau)^2$ in the leap-frog
integration, for constant trajectory length $N\Delta\tau$.
\end{enumerate} 

In addition, 
we compared 
simulation results for the full theory at $(k,y)$, with the output of
a quenched simulation at the corresponding effective coupling value
obtained in a large-$y$ expansion,
\begin{equation}
k^{{\mathrm {eff}}}=k+\frac{2}{y^2}+O\left(\frac{1}{y^4}\right)
\label{keffagain}
\end{equation}
(cf.\ Eq.\ (\ref{kRenorm})).
In table (\ref{TABLATEST}), we give the mean value of several operators
as obtained on a $4^3$ lattice at $k=0.693$, $y=10.0$ and in the
quenched theory. The agreement is excellent. 
Notice that even if the shift in 
the effective coupling is only $3\%$, the effects of the dynamical fermions 
can be clearly measured as  the observables change quite 
significantly  at the critical point $k_{\mathrm c}=0.693$.

\begin{table}[b]
\caption{Comparison of observables in the full
theory (\ref{ACCION}) at $(k = 0.693,y = 10.0)$
and in the quenched model both at the corresponding value of
$k^{{\mathrm {eff}}}$ and at $k_c = 0.693$. 
We have $140,000$ unquenched trajectories ($N$=10, $\Delta\tau$=0.3)
on a $4^3$ lattice. The Metropolis acceptance rate was  65-70\%,
with an autocorrelation time of 3-4 trajectories.}
\medskip
\hrule\hrule
\begin{tabular*}{\hsize}{@{\extracolsep{\fill}}lllll}
Couplings & \multicolumn{1}{c}{$\langle E \rangle$}
& \multicolumn{1}{c}{$\partial_k\langle E \rangle$}
&\multicolumn{1}{c}{$\chi/V$}&\multicolumn{1}{c}{$\xi$}\\\hline
$k$=0.693 , $y$=10.0    & 0.4164(6) &  1.134(6)  & 0.3111(7)   & 2.378(4)\\
$k$=0.713 , $y$=0       & 0.41584(14) &  1.130(4)  & 0.3108(2)   & 2.3779(18)\\
$k$=0.693 , $y$=0       & 0.3928(3) &  1.174(4)  & 0.2836(4)   & 2.214(2)
\end{tabular*}
\label{TABLATEST}
\hrule\hrule
\end{table}
\subsection{Phase Diagram}{\label{NUMERICAL}}

\begin{figure}[htb]
\begin{center}
\leavevmode
\centering\epsfig{file=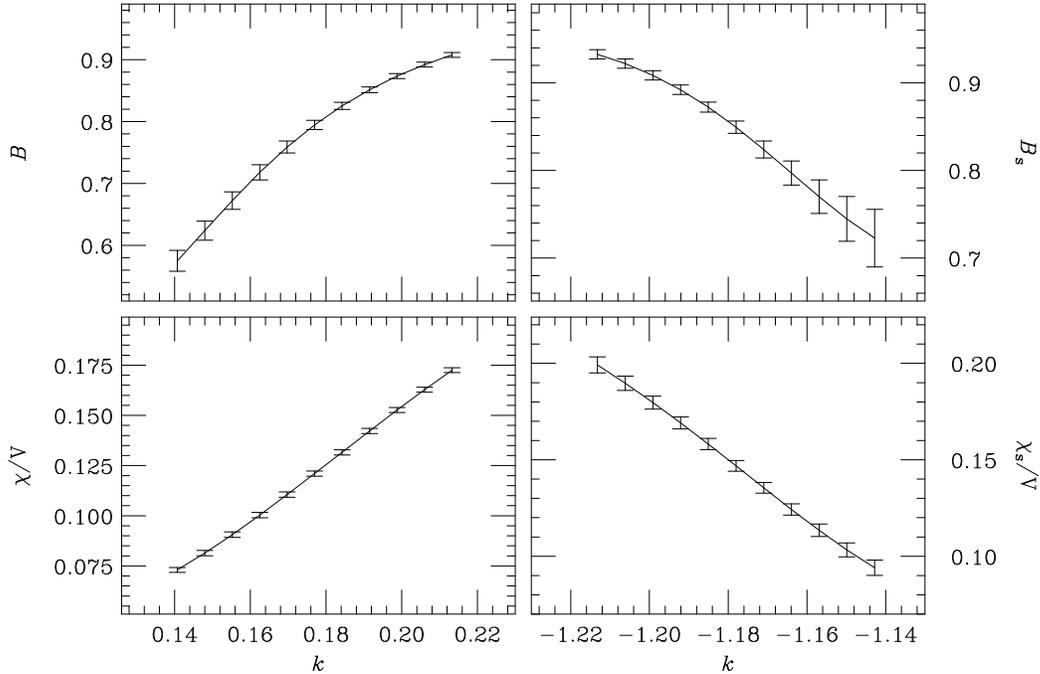,width=0.6\linewidth,angle=90}
\end{center}
\caption{
Binder cumulant (\protect\ref{Bdef}) and non-connected susceptibility
(\protect\ref{susc}) as a function of $k$, around the two critical
points at $y=2.0$. For each critical point, only one simulation
has been carried out. The other points are obtained with the 
Ferrenberg-Swendsen extrapolation method.}
\label{GRANY}
\end{figure}

\begin{figure}[htb]
\begin{center}
\leavevmode
\centering\epsfig{file=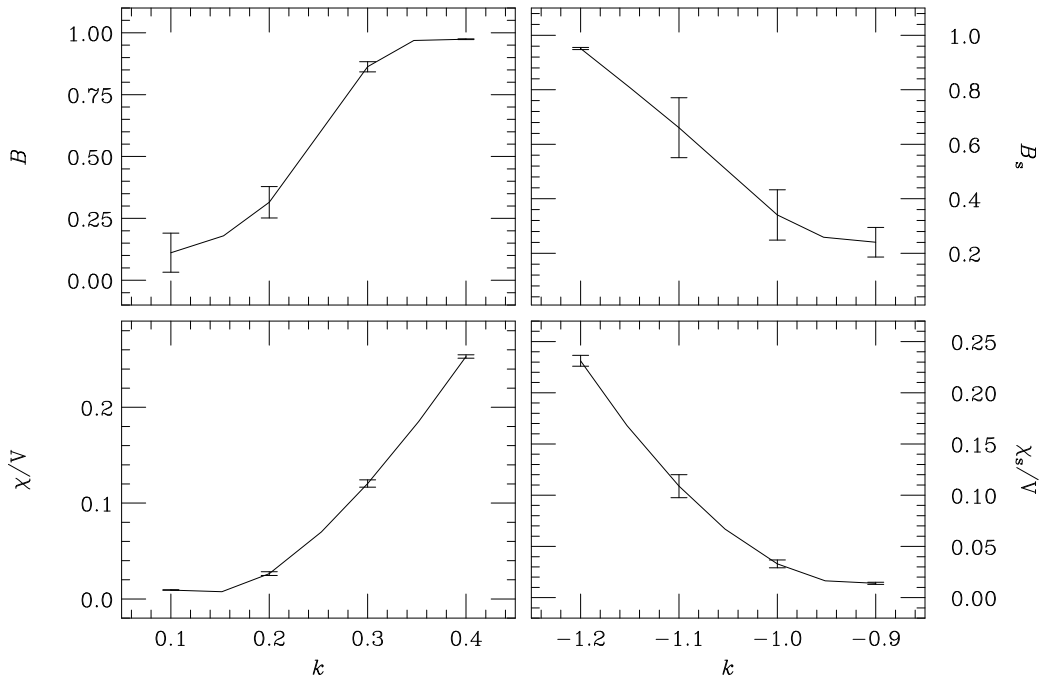,width=0.6\linewidth,angle=90}
\end{center}
\caption{
Binder cumulant (\protect\ref{Bdef}) and non-connected susceptibility
(\protect\ref{susc}) as a function of $k$, around the two critical
points at $y=0.5$. The data points are from different simulations.
}
\label{SMALLY}
\end{figure}

\begin{figure}[htb]
\begin{center}
\leavevmode
\centering\epsfig{file=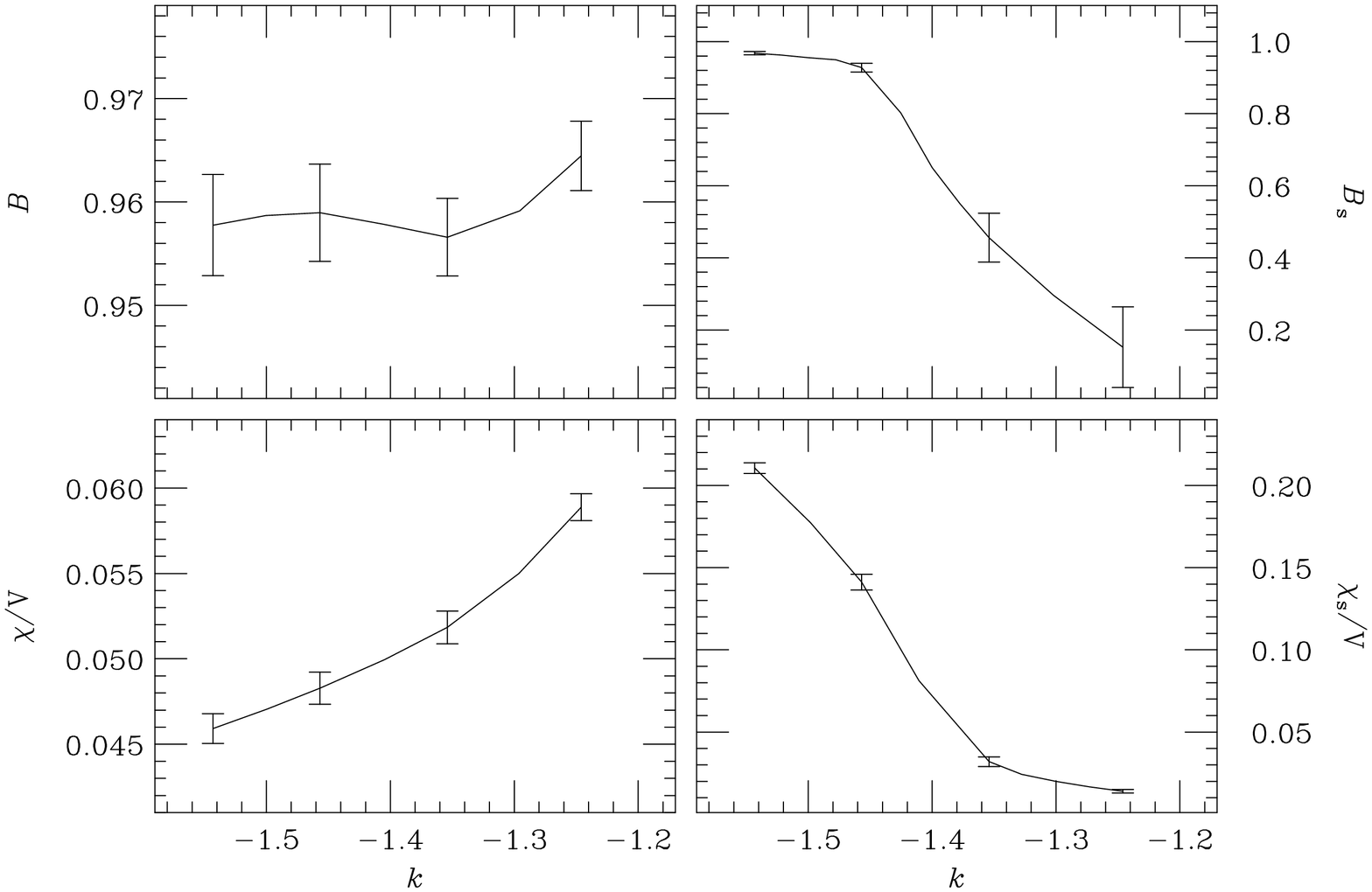,width=0.6\linewidth,angle=90}
\end{center}
\caption{
Binder cumulants and susceptibilities when crossing the FM(S)-EM
transition line at $y=1.15$.
}
\label{EMV}
\end{figure}

\begin{figure}[htb]
\begin{center}
\leavevmode
\centering\epsfig{file=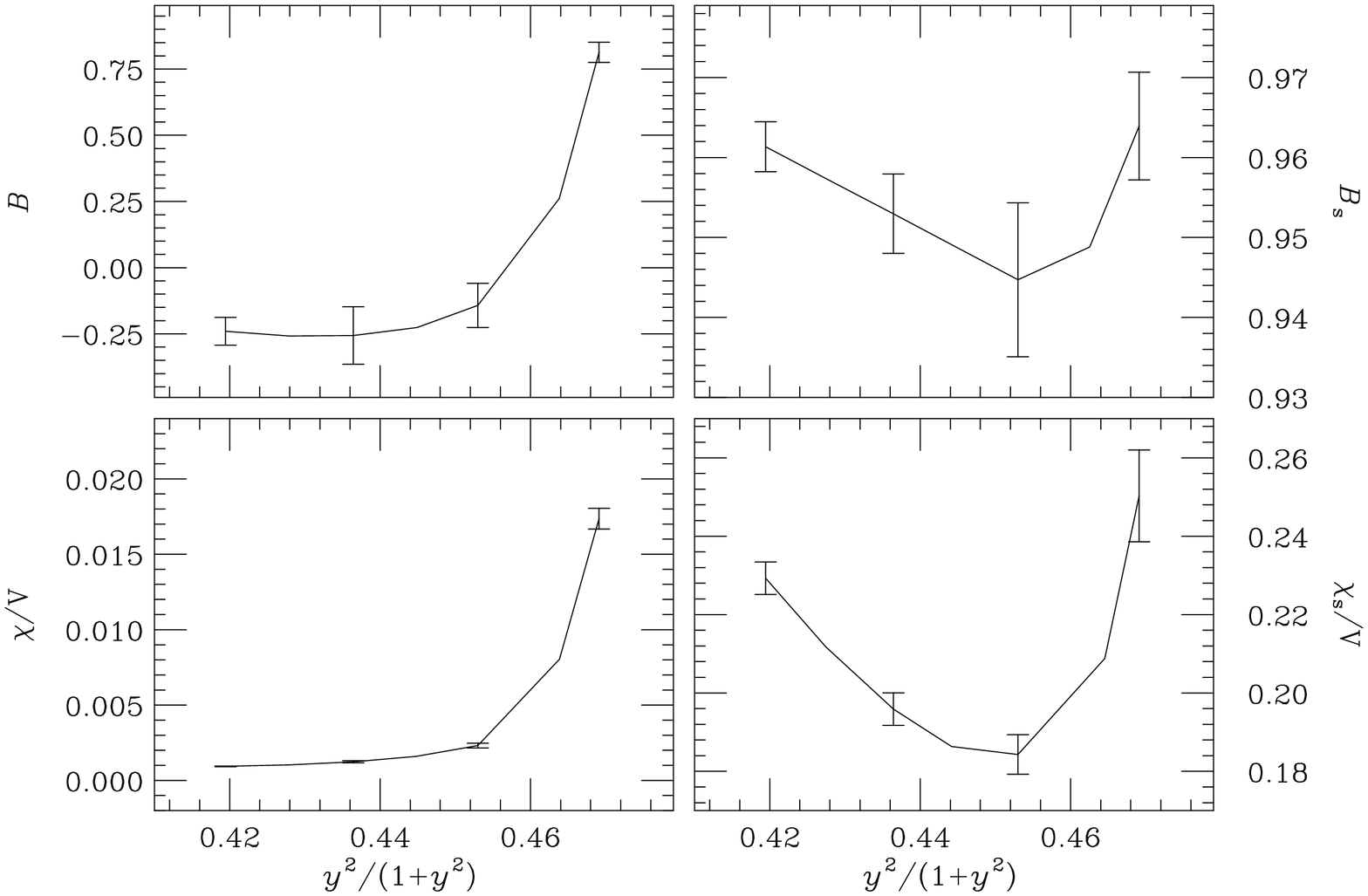,width=0.6\linewidth,angle=90}
\end{center}
\caption{
Binder cumulants and susceptibilities when crossing the AFM(W)-EM
transition line at $k=-1.6$.
}
\label{EMH}
\end{figure}

The phase diagram in fig.\ \ref{PHASES}  was obtained on an $8^3$ lattice.
As there is no true phase transition on a finite lattice, a criterium
is needed to locate the phase boundaries. We looked for the point
where the relevant Binder cumulant equals the value $B=0.8$ it has at
($k=\pm 0.693 \approx k_\mathrm{c}$, $y=0$).
Since $B=1$ deep in the broken phase and
$B\propto 1/L^3$ in the symmetric one, this provides a clean
quantitative criterium which yields a point definitely inside the
critical region. The width of the critical region decreases as
$L^{-1/\nu}$, therefore the systematic error in the critical coupling
will be at most of order $10^{-1}$. However, since the Binder parameter is a
universal quantity, which should stay constant along much of the critical
lines, the error rather goes as $L^{-(\omega+1/\nu)}$ ({\em i.e.} 
${\cal O} (10^{-2})$). Thus, this systematic error is under control in the
full theory as well.  
We used the Ferrenberg-Swendsen extrapolation method to
determine the precise location of the points where $B=0.8$.

The total simulation time was 16 days of the 32 Pentium Pro processor parallel
computer RTNN based in Zaragoza. To allow for a correct
thermalization, we discarded $100$ integrated autocorrelation times of the
relevant susceptibility. This may look utterly conservative, and the
MC history indeed seems to stabilize long before that. However, not
much is known about the {\it exponential} autocorrelation time
of fermionic algorithms and one should be cautious.

As Eq.\ (\ref{STRONG}) shows, both at $y=\infty$ and at $y=0$ we
recover the non-linear $\sigma$-model with its well known
paramagnetic, ferromagnetic and antiferromagnetic phases. At finite
$y$, we expect these phases to extend into the ($k$,$y$) plane.
In fact one can quite precisely anticipate the critical coupling from the
strong coupling formula (\ref{keffagain})
and the quenched critical points $k_{\mathrm c}^{(y=\infty)}=\pm 0.693$.
Using the Ferrenberg-Swendsen extrapolation procedure, the phase
transition lines can be determined down to $y\approx 2.0$.
In fig.\ \ref{GRANY} the variation of the
Binder cumulant and the susceptibility around the two critical
couplings is shown for $y=2.0$.

In the small-$y$ region, the effective action up to ${\cal O}(y^2)$ does
not only renormalize $k$, but also introduces additional couplings,
due to the non-locality of the matrix $K^{-1}$ occurring in the weak-coupling
expansion.
Therefore, we do not have an estimate for $k^{{\mathrm {eff}}}$ as reliable
as in the large-$y$ region (\ref{keffagain}), but we can nevertheless
obtain an estimate for $k_{\mathrm c}(y)$ from the MF approximation.
We have simulated at several values of
the coupling $k$, for fixed $y$, until the corresponding 
Binder parameter crossed its critical value.
A more accurate result for the critical point was later on obtained with the
Ferrenberg-Swendsen extrapolation. In fig.\ \ref{SMALLY},
we have plotted the relevant Binder parameter and susceptibility for
$k$ values near the two critical couplings with $y=0.5$.

In fig.\ \ref{EMV} we show
the variation of both order parameters and Binder cumulants
when crossing the FM(S)-EM transition line at $y=1.15$. We
find a strong change in the staggered quantities, while
the non-staggered ones show a smoother evolution. However,
the non-staggered order parameter is much smaller than its
staggered counterpart. This may indicate that, although
the non-staggered sector is non-critical ($B\sim 1$),
it will eventually undergo a phase transition at lower $k$.
A similar behaviour is found when traversing the AFM(W)-EM line at
$k=-1.6$ (see fig.\ \ref{EMH}), but now the non-staggered quantities
show a more pronounced signal. The detailed study of these transition
lines (order of the phase transitions, critical exponents, etc.) 
requires a finite-size scaling analysis, which is left for future work. 
This study will be much easier if the transition line is crossed
varying $k$, as we lack an analogue of the Ferrenberg-Swendsen extrapolation
method for $y$.

\section{Quasiparticle excitations at the MF level}

\label{sectMFexc}

In this section we explore the relevant excitations involving
fermions, with emphasis on the strong-coupling region of our model.
This will then enable us to discuss electronic properties.

The small-$y$ regime has been studied in relation with
the mechanism by which leptons and quarks acquire
their mass through symmetry breaking in the Electroweak sector of the
Standard Model. Due to the weak coupling, one has essentially
Fermi liquid behaviour, and there are no surprises.
This situation will change dramatically when we consider the strong-coupling
region, though.

\subsection{Fermionic excitations in the FM(S) and PMS phases}

\label{sectPMSexc}

At very large $y$, it is natural to attempt a large-$y$ expansion.
This can be achieved after carrying out the following change of variables:
\begin{eqnarray}
\bar\psi'&=&\bar\psi \, , \label{chofvar1} \\
\psi'&=&\left({\mbox{\boldmath $\phi$}}
       \cdot{\mbox{\boldmath $\tau$}}\right)\psi 
    \label{chofvar2} \, .
\end{eqnarray}
Because of the constraint ${\mbox{\boldmath $\phi$}}^2=1$ and the
identity
$\left({\mbox{\boldmath $\phi$}}\cdot{\mbox{\boldmath $\tau$}}\right)^2=
{\mbox{\boldmath $\phi$}}^2{\mbox{\boldmath $1$}}$, this transformation
has unit Jacobian and its inverse satisfies
\begin{equation}
\psi=\left({\mbox{\boldmath $\phi$}}\cdot{\mbox{\boldmath $\tau$}}\right)\psi'
  \, . \label{chofvarinv}
\end{equation}
In terms of the new variables (dropping the primes) the action takes the form
\begin{equation}
S= -k\, \sum_{x,\mu}\, {\mbox{\boldmath $\phi$}}_x
   \cdot{\mbox{\boldmath $\phi$}}_{x+\mu}+\sum_{x,y}\,
   \bar\psi_x\left(K_{xy}\left({\mbox{\boldmath $\phi$}}_y
   \cdot{\mbox{\boldmath $\tau$}}\right) +y\delta_{xy}\right)\psi_y,
\label{Sagain}
\end{equation}
where the fermion kinetic term is the usual lattice kinetic Dirac operator,
defined in Eq.\ (\ref{MATRIZK}).
After a further rescaling of the $\psi$ fields, the coupling $y$ can be
moved to the kinetic term, where it appears as $1/y$.

Note that this change of variables (\ref{chofvar1},\ref{chofvar2}) was
implicitly present in the MF calculations of the phase diagram in the
strong-coupling region as well (cf.\ Eqs.\ (\ref{Mylarge0},\ref{Mylarge})).

The fermion propagator $\langle \psi_x \bar\psi_y \rangle$ is given by
the expectation value of the inverse fermion matrix,
which in a large-$y$ expansion becomes
\begin{equation}
\langle \psi_x \bar\psi_y \rangle \ = \ \left\langle M^{-1}_{xy}\right\rangle
 \ = \ \left\langle \frac1y \left(1 - \frac1y K ({\mbox{\boldmath
 $\phi\cdot\tau$}})+ \frac{1}{y^2} K ({\mbox{\boldmath
 $\phi\cdot\tau$}}) 
K ({\mbox{\boldmath $\phi\cdot\tau$}}) - \ldots
  \right)_{xy}\right\rangle
   \, .  \label{fermprop}
\end{equation}
This can be viewed as a sum over paths of increasing length connecting
$x$ and $y$ (recall that $K$ is a nearest-neighbour matrix).

In the FM(S) phase, there is a non-zero magnetization
$v = |\langle {\mbox{\boldmath $\phi$}} \rangle|$.
Expectation values of products of ${\mbox{\boldmath $\phi$}}$ fields on different sites
are replaced by the appropriate powers of $v$.
Corrections to this approximation as well as contributions from paths
visiting a given site more than once are of higher order in $1/d$ and
are ignored at the MF level.
The series (\ref{fermprop}) can thus be resummed and one finds a
propagator
\begin{equation}
\langle \psi \bar\psi \rangle_{FM(S)} \ = \ 
\frac{1/v}{K + y/v}
\end{equation}
which is that of a fermion with a mass $y/v$.
Note that, since $v<1$, this is a huge mass if $y$ is large.
The propagator for the original fermion, before the change of variables
(\ref{chofvar1},\ref{chofvar2}), corresponds to the same physical particle;
the only difference is in the wave function renormalization.

In the PM(S) phase, $v=0$, so at the MF level the fermion would be
infinitely massive, or in other words, non-propagating.
Beyond this naive MF level, however, a large but finite mass will be found.
This is due to the next-to-leading contributions to the series
(\ref{fermprop}).
The dominant terms are now those involving the expectation value for
the nearest-neighbour energy $z^2 \equiv \langle {\mbox{\boldmath
$\phi$}}_x \cdot {\mbox{\boldmath $\phi$}}_{x+\hat\mu}
\rangle$, which is of order $1/2d$ and therefore absent at the MF level.
The resummation of contributions in (\ref{fermprop}) now leads to a
fermion propagator with a mass $y/z$, which is even larger than 
the mass of the fermion in the FM(S) phase.

The conclusion of this analysis, which is similar to that in (chiral) Yukawa
models in the Electroweak theory \cite{SMITz2}, is that the elementary
fermion excitations in the large-$y$ region are very heavy (hence essentially
non-propagating), and therefore
play no role in the spectrum of light excitations.
This holds even stronger in the PMS phase than in the FM(S) phase.

\subsection{Fermionic excitations in the AFM(S) phase}

\label{AFMSsection}

Here our point of departure is again the form of the action (\ref{Sagain}),
which is tailored for studying the large-$y$ behaviour.
In the AFM(S) phase, we have a staggered expectation value for the
${\mbox{\boldmath $\phi$}}$ field at the MF level, which can be taken in the 3-direction,
\begin{equation}
{\mbox{\boldmath $\phi$}}_x=\epsilon_x v\left(\begin{array}{c}
0\\0\\1\end{array}\right)
\label{MSTAG}
\end{equation}
(with $\epsilon_x=(-1)^{x_1+x_2+x_3}$).
Hence
\begin{equation}
\left({\mbox{\boldmath $\phi$}}_x\cdot{\mbox{\boldmath $\tau$}}\right)\psi_x=
\left(\begin{array}{r} v\,\epsilon_x\,\psi_x^{(1)}\\
-v\,\epsilon_x\,\psi_x^{(2)}\end{array}\right),
\end{equation}
where $\psi_x^{(i)}$, $i=1,2$ labels the two {\it flavours} in $\psi_x$. So
after the change of variables (\ref{chofvar1},\ref{chofvar2})
the kinetic operator in (\ref{Sagain}) is still diagonal in flavour.
The only effect of the new variables is to change the lattice Dirac
operator from (\ref{MATRIZK}) to
$$ v\,\epsilon_y\, \tau_3 K_{xy}.$$
Due to this diagonal structure in flavour space, we can concentrate on
one flavour, say $\psi^{(1)}$; the other flavour is obtained by taking $-v$
instead of $v$.
In Fourier space, the kinetic term for $\psi^{(1)}$ is given by
\begin{equation}
   -i\,v\,\ 
      {\slash\hspace{-0.8em}{\sin}}\, p\ \delta_{p,q\pm \pi},
\end{equation}
where
\begin{eqnarray}
{\slash\hspace{-0.8em}{\sin}}\, p&=&\sum_{\mu}\,\sigma_{\mu}\ \sin\,p_{\mu}
 \, , \label{sinp} \\
 \delta_{p,q\pm \pi}&=&\prod_\mu \delta_{p_\mu,q_\mu+\pi \ \mathrm{mod}\ 2\pi}
   \, . \label{deltapq}
\end{eqnarray}
So we obtain for the inverse of the MF propagator
in the AFM(S) phase, 
\begin{equation}
M_{p,q}=-i\, v\ \,{\slash\hspace{-0.8em}{\sin}}\, p\ \delta_{p,q\pm \pi}+
y\,\delta_{p,q},
\label{PROAFMRAW}
\end{equation}
or, in matrix notation for the subspace of the modes coupled in 
Eq.\ (\ref{PROAFMRAW}), $p$ and $p\pm(\pi,\pi,\pi)$,
\begin{equation}
M_{p,p\pm(\pi,\pi,\pi)}=\left(\begin{array}{cc} 
y & -i\, v\  {\slash\hspace{-0.8em}{\sin}} p\\
i\, v\  {\slash\hspace{-0.8em}{\sin}} p & y
\end{array}\right).
\label{PROAFMMAT}
\end{equation}

To find the quasiparticle excitations in the AFM(S) phase we diagonalize
the fermionic part of the action (\ref{PROAFMMAT}).
One obtains
\begin{equation}
S=\int_p \bar\psi(p) \,
(y-v\ \; {\slash\hspace{-0.8em}{\sin}}\, p) \, \psi(p)\ ,
\label{SDIAG}
\end{equation}
where
$$\psi(p)=\frac{1}{\sqrt{2}}\left[\psi^{(1)}(p)\,+
\, i\,\psi^{(1)}(p+\pi)\right],$$
or, in {\em position} space,
$$\psi_x=\frac{1}{\sqrt{2}}\left[\psi^{(1)}_x\,+
\, i\,\epsilon_x\psi^{(1)}_x\right].$$

The momentum space propagator corresponding to (\ref{SDIAG}) is thus
\begin{equation}
S(p)=\frac{1}{y-v\ \; {\slash\hspace{-0.8em}{\sin}}\, p}=
\frac{y+v\ \; {\slash\hspace{-0.8em}{\sin}}\, p}
{y^2-v^2\,\sum_{\lambda} \sin^2\, p_{\lambda}}.
\label{PROAFM}
\end{equation}
Since we are working in imaginary time, one would expect quasiparticle poles
in $S(p)$ to appear at negative values of $p^2$.
The unusual relative minus sign in the denominator (\ref{PROAFM}) therefore
does not seem to allow for a quasiparticle interpretation, at first sight.

However, (\ref{PROAFM}) suggests the possibility of
light excitations with a relativistic dispersion relation around spatial
momenta $\left(\pm\frac{\pi}{2},\pm\frac{\pi}{2}\right)$.
To see this, consider the denominator in Eq.\ (\ref{PROAFM})
for small $k_\mu = p_\mu \pm \pi/2$:
\begin{equation}
y^2-v^2\,\sum_{\lambda} \sin^2\, p_{\lambda}\ =\ (y^2-v^2\, d)\ +\ 
v^2\,\sum_{\lambda}\,k_{\lambda}^2\ +\ {\cal O}(k^4),
\label{TAYLOR}
\end{equation}
where $d=3$ is the space-time dimension.
As long as we are at large enough $y$, such that $y^2\,>\, d\,v^2$
(recall $v^2<1$), this dispersion relation corresponds to a
relativistic excitation with $m^2=(y^2-d v^2)/v^2$, in this naive
MF calculation.
Several comments are in order:

\begin{enumerate}
\item For $v$=0, we recover the MF result for the PMS phase:
the kinetic term in (\ref{TAYLOR}) is suppressed.
\item At the MF level, only for $(y^2-dv^2)$ small enough compared to
$v^2$ these fermionic excitations,
$\left({\mbox{\boldmath $\tau$}}\cdot{\mbox{\boldmath $\phi$}}\right)\psi$,
can propagate easily. Since $v^2<1$, this can only happen for $y^2$
not too large.
Whether or not this situation will arise depends on the precise
phenomenological relation between $x$ and $(y,k)$, which is likely 
to be dopant-dependent (see section \ref{PHEN}).
\item These would-be excitations are characteristic of the AFM(S) phase.
Let us recall that in the PMS phase no {\it light} fermionic excitations have 
been identified at the MF level.
In any case, in the SC phase of our model {\it 
we are not in an AFM background}. Therefore no momentum-space rotation
will be needed to interpret the fermionic excitations and no
hole-pockets around $\left(\pm\pi/2,\pm\pi/2\right)$ are expected in the SC
phase~\cite{NORMAN}.

\item We need to fix the scale in some way, as done for instance 
in~\cite{CHAKRA}, to estimate by means of a MC calculation the masses of
all possible excitations of our model.

\end{enumerate}

\subsection{Light bound states in the PMS phase}

\label{excPMSbos}

\begin{figure}[htb]
\begin{center}
\leavevmode
\centering\epsfig{file=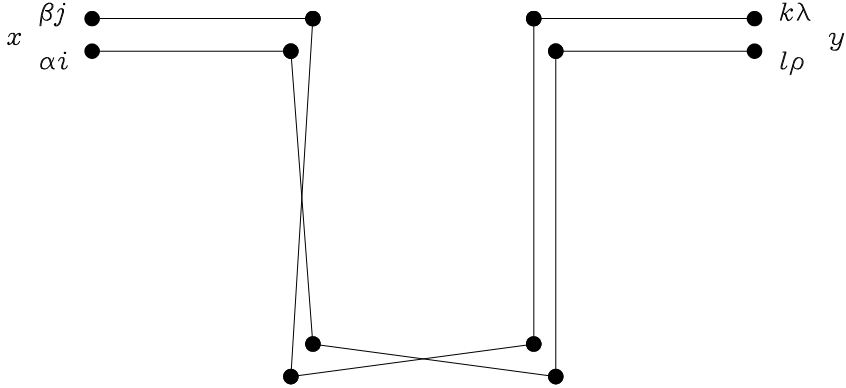,width=0.35\linewidth,angle=90}
\end{center}
\caption{
A typical double-chain diagram, connecting sites $x$ and $y$.
The chains are parallel in position space.
}
\label{DOUBLECHAIN}
\end{figure}

We have seen above that the fermionic excitations in the PMS phase are
very heavy.
We will now show that there are bound states of elementary fermions,
however, which are light.
This is done by means of a MF calculation of the double-chain type
\cite{STEPHANOV}.
These pairs will be the electron or hole pairs responsible for
the superconductivity, as we will discuss in Sect.\ \ref{phasediagcup}.

Consider the propagator for a fermion pair $\psi_x\psi_x$,
\begin{equation}
 \langle \psi^\alpha_{x,i} \psi^\beta_{x,j}
     \bar\psi^\lambda_{y,k} \bar\psi^\rho_{y,l} \rangle
\ = \
\langle M^{-1}_{x,\beta,j;\,y,\lambda,k} \, M^{-1}_{x,\alpha,i;\,y,\rho,l} \rangle -
\langle M^{-1}_{x,\alpha,i;\,y,\lambda,k} \,
     M^{-1}_{x,\beta,j;\,y,\rho,l} \rangle\ .
\label{pairprop}
\end{equation}
Here $M^{-1}$ is the single-fermion propagator,
$\alpha, \beta, \lambda, \rho$ are Dirac indices, and $i, j, k, l$
are flavour indices. Thus, this propagator is really a $16\times 16$ matrix.
For the moment we keep all these indices as they are; later on we will discuss
how pairs of them decompose into quantum numbers for the composite state.

Let us concentrate on the first $\langle M^{-1} M^{-1} \rangle$ term in
Eq.\ (\ref{pairprop}).
Using the $1/y$ expansion of $M^{-1}$ as before, we find the series
\begin{equation}
\langle M^{-1}_{x,\beta,j;\,y,\lambda,k} \, M^{-1}_{x,\alpha,i;\,y,\rho,l} \rangle
 \ = \ 
\sum_{N,N'=0}^\infty \left\langle \left[ \frac{\phi}{y} \left( K\frac{\phi}{y}
  \right)^N \right]_{x,\beta,j;\,y,\lambda,k}
 \left[ \frac{\phi}{y} \left( K\frac{\phi}{y}
  \right)^{N'} \right]_{x,\alpha,i;\,y,\rho,l} \right\rangle
 \, , \label{pairseries}
\end{equation}
where we have written $\phi$ as a shorthand for 
$({\mbox{\boldmath $\phi$}}\cdot{\mbox{\boldmath $\tau$}})$. It is clear
that 
only terms with $N+N'$ even survive in a paramagnetic phase, due
to the ${\mbox{\boldmath $\phi$}}\rightarrow -{\mbox{\boldmath $\phi$}}$ symmetry, thus a factor $(-1)^{N+N'}$
has been dropped.
Since the matrix $K$ connects nearest-neighbour sites only, each term
in this series can be seen to represent a product of two paths (chains) of
lengths $N$ and $N'$ respectively, connecting site $x$ with site $y$
(so, if the ``distance'' between $x$  and $y$ is even(odd), both $N$ and
$N'$ will be even(odd)).

We will attempt to sum the complete series, to leading order in $1/d$,
where $d=1+2=3$ is the euclidean space-time dimension. For this, we
need the spin-spin propagator, which in this approximation
is extremely short ranged
\begin{equation}
\langle \phi_x^a \phi_{x}^b \rangle = \frac{1}{3} \delta^{ab}
 \, .
\label{SHORT}
\end{equation}
Expectation values of the type $\langle {\mbox{\boldmath $\phi$}}_x \cdot {\mbox{\boldmath $\phi$}}_{x+\hat\mu} \rangle$
are of order $1/d$, and others are suppressed even stronger.
Thus, assuming (\ref{SHORT}),
we observe that any term in the series which contains ${\mbox{\boldmath $\phi$}}_x$ for
a given site $x$ only once or an odd number of times will vanish due to
$\langle {\mbox{\boldmath $\phi$}} \rangle = 0$.
When the site is visited twice, it follows from ${\mbox{\boldmath $\phi$}}^2 = 1$ that
the contribution from the ${\mbox{\boldmath $\phi$}}$ fields
is proportional to $\frac{1}{3} \delta^{ab}$.
Thus each site along the chains connecting $x$ and $y$ must be visited
an even number of times.
One class of diagrams fulfilling this requirement consists of the
so-called `double-chain' diagrams, where the propagation of both fermions
between $x$ and $y$
follows the same path in position space (see figure \ref{DOUBLECHAIN}).
As was convincingly argued in Ref.\ \cite{STEPHANOV}, this class
saturates the dominant diagrams in the $1/d$ expansion.
Indeed, one can easily check by concrete examples, how deviations from
double-chain behaviour induce additional powers of $1/d$.
We shall also assume that these double chains are self-avoiding (this
is allowed at first order in $1/d$).

\begin{figure}[htb]
\begin{center}
\leavevmode
\centering\epsfig{file=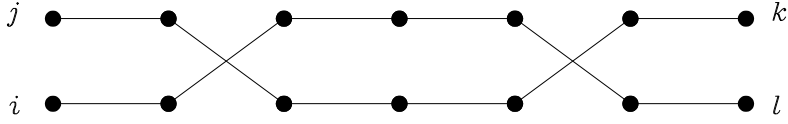,width=0.11\linewidth,angle=90}
\end{center}
\caption{
A matrix-product term contributing to the flavour structure.
}
\label{FLAVOUR}
\end{figure}

Our task is thus to sum up all double chain diagrams connecting $x$ and $y$.
Let us first consider the flavour structure.  
Using  (\ref{SHORT}) one finds that
\begin{equation}
\langle ({\mbox{\boldmath $\phi$}}_x\cdot{\mbox{\boldmath $\tau$}})_{jk} ({\mbox{\boldmath $\phi$}}_x\cdot{\mbox{\boldmath $\tau$}})_{il} \rangle
\ = \ \frac13 \sum_a \tau^a_{jk} \tau^a_{il}
\ = \ \frac13 (- \delta_{jk} \delta_{il} + 2 \delta_{jl} \delta_{ik}) 
 \, . \label{phitauphitau}
\end{equation}
From this and from the ultra-local correlations
we are considering (cf.\  Eq.\  (\ref{SHORT})), it follows that
the product of $2(N+1)$ factors of $({\mbox{\boldmath $\phi$}}\cdot{\mbox{\boldmath $\tau$}})$
along a double chain of length $N$ visiting the points $x=x_0, x_1,\ldots,
y=x_N$ (cf.\ Eq.\ (\ref{pairseries})) is
\begin{equation}
\left\langle
   \left[
     \prod_{n=0}^N ({\mbox{\boldmath $\phi$}}_{x_n}\cdot{\mbox{\boldmath $\tau$}}) 
   \right]_{x,j;\,y,k}
  \,
   \left[
     \prod_{n'=0}^N ({\mbox{\boldmath $\phi$}}_{x_{n'}}\cdot{\mbox{\boldmath $\tau$}}) 
   \right]_{x,i;\,y,l}
\right\rangle
\ = \
P\,\delta_{jk} \delta_{il}\  + \ Q\,\delta_{jl} \delta_{ik} . 
\end{equation}
To calculate $P$ and $Q$, it is convenient to represent the general term 
contributing to the above matrix product as in figure \ref{FLAVOUR}. A
graph contributing to $\delta_{jk} \delta_{il}$ will have an even number
of crossings, while diagrams contributing to $\delta_{jl} \delta_{ik}$
jump an odd number of times. Each crossing contributes a factor $\frac{2}{3}$,
while non-crossings yield factors $-\frac{1}{3}$ (cf.\  Eq.\  (\ref{phitauphitau})).
Now, $P$ and $Q$ can be easily obtained using binomial summation:
\begin{eqnarray}
&&\left\langle
   \left[
     \prod_{n=0}^N ({\mbox{\boldmath $\phi$}}_{x_n}\cdot{\mbox{\boldmath $\tau$}}) 
   \right]_{x,j;\,y,k}
  \,
   \left[
     \prod_{n'=0}^N ({\mbox{\boldmath $\phi$}}_{x_{n'}}\cdot{\mbox{\boldmath $\tau$}}) 
   \right]_{x,i;\,y,l}
\right\rangle
  \nonumber \\
&&\ \ \ \ \ \ \ \ \ \ \ \ \ \ \  \ = \
\left( \frac13 \right)^N \,
      \frac12 ( \delta_{jk} \delta_{il} + \delta_{jl} \delta_{ik} )
 \ + \
(-1)^N \, \frac12 ( \delta_{jk} \delta_{il} - \delta_{jl} \delta_{ik} )
 \, , \label{phitauN}
\end{eqnarray}
where we have separated in a term symmetric under
$(ji)\leftrightarrow (ij)$ and an antisymmetric one (this will be needed
for separating the contribution to different quantum numbers).
It is remarkable that the flavour contribution only depends on the
double-chain length, but not on its shape. This allows for a total
factorization between flavour and Dirac indices.

Next, consider the Dirac structure.
One gets products of matrices
\begin{equation}
K^{\mu_n}_{x_n x_{n+1}} K^{\mu_n}_{x_n x_{n+1}}
\sigma^{\mu_n}_{\beta_n\lambda_n}   \sigma^{\mu_n}_{\beta_{n+1} \lambda_{n+1}},
 \label{KKss}
\end{equation}
along the double chain,
where 
\begin{equation}
K^\mu_{xy} \ = \ 
  \frac12 (\delta_{y,x+\hat\mu} - \delta_{y,x-\hat\mu})
 \, . \label{Kagain}
\end{equation}
One readily finds that
\begin{equation}
  A_{xy}^\mu\equiv 4\, K^\mu_{xy}K^\mu_{xy} \ = \ (\delta_{y,x+\hat\mu} \ + \ \delta_{y,x-\hat\mu})
 \, . \label{Amudef}
\end{equation}
Thus, we need to calculate
\begin{equation}
\sum_{\{\mu_n\}} \left[\prod_n\ {\frac{1}{4}} A^{\mu_n}\, 
\sigma^{\mu_n}\otimes\sigma^{\mu_n}\right]_{x,\beta , \alpha ; y, \lambda ,\rho
}\ ,
\label{AMUNN}
\end{equation}
where the sum is extended to all the lattice paths (denoted by $\{\mu_n\}$)  of-length $N$ connecting
$x$ and $y$. Now, we can extend the sum to {\it all} length-$N$
lattice paths starting at $x$, 
because paths not connecting $x$ to $y$ will contribute a zero $x\, y$ entry. 
This can be also understood   by realizing
that once the chain has arrived at $x_i$, there are $2d$ possible
directions to continue the chain.
These are added up by summing Eq.\ (\ref{KKss}) over $\mu$.
At the next site, we do the same for the next step along the chain.
The contributions of all double chains are therefore added up when
we take the product of these $\mu$-sums along the chain.
Corrections due to backtracking ($2d \rightarrow 2d-1$) are down by $1/d$.

So we need to calculate powers of the matrix
\begin{equation}
\frac{1}{4}
\sum_\mu A^\mu\, 
\sigma^\mu\otimes\sigma^\mu
  \, . \label{sumKKss}
\end{equation}
One way to do that is to write it out explicitly as a $4\,\times\,4$ matrix in
the space spanned by the vectors $(\beta,\lambda) = $(1,1), (2,2), (1,2)
and (2,1), in that order.
One finds that it equals
\begin{equation}
    \frac{1}{4}
    \left(
  \begin{array}{cccc}
       A^3 & A^1 - A^2 & 0 & 0 \\
      A^1 - A^2 &  A^3 & 0 & 0 \\
      0 & 0 & -A^3 & A^1 + A^2 \\
      0 & 0 & A^1 + A^2 & -A^3 
  \end{array}
    \right)\ .
\label{Amatrix}
\end{equation}
It can be diagonalized in this $4\,\times\,4$ space.
The eigenvalues, up to the factor $1/4$, are found to be
\begin{eqnarray}
&&\lambda^{\mu} \ = \ A \ - \ 2 A^{\mu} \ \ \ \ \ \ \ \ \ \ \ \ (\mu=1,2,3)
  \label{lambda123}\, , \\
&&\lambda^4 \ = \ - A\, ,
  \label{lambda4}
\end{eqnarray}
where
\begin{equation}
A \ = \ \sum_{\mu=1}^3 A^\mu  \ = \  \Box + 2d
 \, , \label{Adef}
\end{equation}
and $\Box$ is the lattice discretization of the d'Alembertian
$\sum_\mu \partial_\mu \partial_\mu$.
The $N^\mathrm{th}$ power (see (\ref{AMUNN})) of the matrix (\ref{sumKKss}) is now easy to calculate.

In order to collect the factors and sum up the contributions, let us
go back to Eq.\ (\ref{pairprop}).
We see that we need to antisymmetrize each term in $\langle M^{-1} M^{-1}
\rangle$ with respect to the simultaneous interchange of
$\alpha, i$ with $\beta, j$.
This gives a sum of two terms, one symmetric in $\alpha \leftrightarrow \beta$
and antisymmetric in $i \leftrightarrow j$ (corresponding to a composite
state which is a Dirac vector and a flavour singlet), and one vice versa
(singlet in Dirac space, vector in flavour space).
Note that Eq.\ (\ref{phitauN}) has already been written as a sum of
symmetric and antisymmetric terms.
The symmetric and antisymmetric parts of the Dirac structure correspond
to Eqs.\ (\ref{lambda123}) and (\ref{lambda4}), respectively.

Collecting the various factors, we can carry out the geometric sum over $N$ in
Eq.\ (\ref{pairprop}) and we find
the following propagators for the composite states:
\begin{itemize}
\item
a Dirac vector -- flavour singlet with propagator
\begin{equation}
\frac{8 \delta^{\mu\nu}}{-\Box + 2 A^{\mu} - 4 y^2 - 2d}
\label{propvecsing}
\end{equation}
where $a,b$ are the Dirac vector indices
\item
a Dirac singlet -- flavour vector with propagator
\begin{equation}
\frac{-8 \delta_{IJ}}{-\Box - 12 y^2 - 2d}
\label{propsingvec}
\end{equation}
where $I,J$ are the flavour vector indices.
\end{itemize}

These have the form of massive bosonic propagators, up to the
following caveat 
(of course, higher order corrections in $1/d$ may induce shifts
in the precise location of the poles, as well as their residues).

The propagators in (\ref{propvecsing}) contain the matrix $2A^{\mu}$ in the
denominator.
However, this term must be ignored since it is sub-dominant in $1/d$, compared
with the (lattice) d'Alembertian $\Box$.

The numerator of the propagator (\ref{propvecsing})
carries a delta function only, instead of the usual tensor structure
$\delta_{\mu\nu} - \partial_\mu \partial_\nu / m^2$.
This is also an artifact of the $1/d$ approximation.

Notice also that the terms which would play the role of a mass squared
in the denominators have an apparently wrong sign.
However, it is easy to check that the composite field $\epsilon_x\psi_x\psi_x$
(where $\epsilon_x = (-1)^{\sum_\mu x_\mu}$ as usual)
does lead to a massive Dirac singlet -- flavour vector propagator with
mass squared $m^2_{(0,1)} = 12 y^2 -2d = 12y^2 - 6$.
Similarly,  one obtains a massive Dirac vector --
flavour singlet with a mass squared $m^2_{(1,0)} = 4 y^2 - 2d = 4 y^2 - 6$.
We thus conclude that the right interpolating field is 
$\epsilon_x\psi_x\psi_x$~\cite{STEPHANOV} (one could equally well consider it
as an excitation centered around spatial momentum $(\pi,\pi)$, though).

The conclusion is that we find massive bound states of fermions in the
PMS phase.  They are bound by the strong interactions with the spin waves.
These composites are lighter
than the elementary fermions in this phase, when $y$ moves away from
the value $\infty$.  This means that they will be
the dominant light excitations, and we will argue that they condense
and lead to superconductivity in our model.

\section{The $x$--$T$ phase diagram of the cuprates}

\label{phasediagcup}

In this section we shall consider the application of the above results
to the cuprates. Strictly speaking, our calculations only apply to
the zero-temperature case. However, some conjectures can be made for the
case of non-zero temperature, by means of very general thermodynamic 
considerations.

\subsection{Physics at zero temperature}

\label{physzeroT}

We will discuss how the copper oxide materials are described by our model,
by following the `trajectory' which they map out in the $y, k$ phase
diagram of our model as the doping fraction $x$ is increased from zero
all the way to the overdoped regime.
Along the way, various phases will be traversed, and the relevant light
excitations present will be given their physical interpretation.

As we discussed earlier on, it is by now well established that
the undoped material can be described by a point
($k\lesssim k_{\mathrm c}$,$y=\infty$) in the AFM(S) phase of the O(3) model
to which our model reduces for $y=\infty$.
The actual value
of $k$ is related to the spin stiffness and velocity; in the large-$S$
approximation $k\propto S$ \cite{CHAKRA}.  As the doping fraction $x$
is increased, the carrier mobility increases which we have argued 
in section \ref{PHEN} to correspond to decreasing $y$ in our model.

We have schematically indicated this `evolution' of the cuprates with
increasing doping by the line of arrows in Fig.\ \ref{PHASES}.
For illustrative purpose, let us assume (see our comment in section \ref{PHEN})
a relation of the type $x\sim C^2/(C^2+y^2)$ for some constant $C$ (in the
case $C=1$ the horizontal axis in Fig.\ \ref{PHASES} would then correspond to
$1-x$), but this is immaterial for the qualitative picture.

Note that we move in the direction of the PMS--AFM(S) transition line,
so the AFM order will decrease.
This is consistent with the experimentally observed reduction of AFM order
upon doping.

At some point along our trajectory, still within the AFM(S) phase,
the possible excitations at the $(\pm\pi/2,\pm\pi/2)$ hole pockets
(cf.\ Sect.\ \ref{AFMSsection}) would become light and start to dominate.

When the doping is increased even more, we move into the PMS phase.
As long as we remain close to the PMS--AFM(S) transition in this phase,
short-range AFM correlations will still be present,
although they are predicted to decrease as one moves deeper into this 
phase. The crucial point is that, as demonstrated in the MF calculation
of Sect.\ \ref{excPMSbos}, the only light excitations left now
are the ``PMS pairs'', bosonic fermion bound states.
Let us remark that a rich spectrum of excitations with different
quantum numbers (not all of them identified in our MF calculation)
should be expected. In our calculation, the flavour singlet
({\em i.e.} the physical spin-singlet) has turned out to be lighter than the
triplet, which is encouraging. To investigate this fundamental
point in more detail, a  MC calculation has to be done.

At temperature $T=0$ these pairs will be Bose-Einstein (BE) condensed, as any
other bosonic state would.
This leads to superfluid behaviour for the pairs, and to
superconductivity once electromagnetism is coupled into the model.
Since a finite number ${\cal O}(x_{c_1}/2)$ of them becomes available at
the same time, at the point where we enter the PMS phase
(corresponding to some critical doping $x_{c_1}$),
one expects a finite ({\em i.e.}, not infinitesimally small) critical
temperature $T_{\mathrm c}=T_{\mathrm {BE}}$ here.

Following the arrow line towards
even smaller $y$, we leave the superconducting PMS phase at $x_{c_2}$.
Thus we understand why
{\it superconductivity only happens for a  range of doping}.
Again, $T_{\mathrm c}$ is expected to remain finite up to $x_{c_2}$.
This may explain why $T_{\mathrm c}$ is experimentally observed to jump
steeply at $x_{c_{1,2}}$; however, we will discuss the
possibility of even more unusual behaviour when we discuss the $T \neq 0$
behaviour below.

Our model predicts a ferromagnetically ordered phase at $T=0$ in this
intermediate-$y$ regime.
The system traverses a region where one needs to go over from
a description of the charge carriers in terms of the variables
$\left({\mbox{\boldmath $\tau$}}\cdot{\mbox{\boldmath $\phi$}}\right)\psi$
at large $y$, to a description in terms of the fields
$\psi$ appropriate for the \hbox{small-$y$} Fermi-liquid regime.
(There may be a kind of cross-over in this region.)
In either description, propagation of the carriers is hampered by
the effective mass induced by the FM spin waves.
 
At this point one may object that no FM behaviour has been observed
experimentally in the overdoped regime.
We note, however, that at some non-zero temperature
(cf.\ Sect.\ \ref{nonzeroT}), this phase will be replaced with a
thermally disordered phase.
Since the overdoped regime has not been explored in as much detail as the
underdoped and superconducting regimes, in particular at very low
temperatures, this may explain why no FM behaviour has been detected here.

For very large doping we reach the weak-coupling region with
Fermi-liquid behaviour.
We will end up in the PMW, AFM(W) phase or possibly remain in the FM(W) phase of
our model.
This will depend on the actual trajectory of the system with doping
(see our comments in section \ref{PHEN}, and recall
that for large doping fraction the effective $k$ parameter might also
be influenced by it).

All these considerations illustrate how such widely varying behaviour in
the cuprates, controlled by the doping fraction $x$, 
is reproduced by varying just one parameter in our simple two-parameter model.
A point needing further investigation is the following. At strong
coupling, one would expect a shift from commensurate to incommensurate AFM ordering,
with increasing doping~\cite{SHRAIMAN}. After integration out the fermion
field in our model, frustrating couplings (of order $1/y^4$ in the strong 
coupling expansion) are generated, which presumably lead to such a phenomenon.
Our MC simulation, was done on a too small lattice to be able to resolve such 
an effect, and our MF is not reliable in this intermediate-$y$ 
region~\cite{class}.

An interesting prediction of our model is that
superconductivity is unlikely to occur in materials with spin $S>1/2$.
The reason is that the undoped model would correspond to a point
$k \ll k_c$ in the O(3) model ($|k|\propto S$, see ref.\ \cite{CHAKRA}),
making it unlikely that the ``evolution trajectory'' 
pass through the PMS phase.
In more physical terms, the larger  $|k|$ is in our model, the stiffer
is the magnetic ordering and the less effective are the  spin 
fluctuations needed to bind the fermions.
For instance, upon doping the layered compounds La$_{1-x}$Sr$_{1+x}$MnO$_4$,
which have localized $S=3/2$ spins (implying $k\sim -2$ in the O(3) model)
a disappearance of 
the antiferromagnetic phase and the subsequent emergence of 
an exotic magnetic phase (maybe a spin-glass phase) is observed
\cite {MORITOMO}, but no superconductivity appears.
Essentially, the same thing happens for insulating nickelates 
La$_{2-x}$Sr$_{x}$NiO$_4$, for which the localized spin is $S=1$~\cite{NICKEL}.

\subsection{Physics at non-zero temperature}

\label{nonzeroT}

\begin{figure}[htb]
\begin{center}
\leavevmode
\centering\epsfig{file=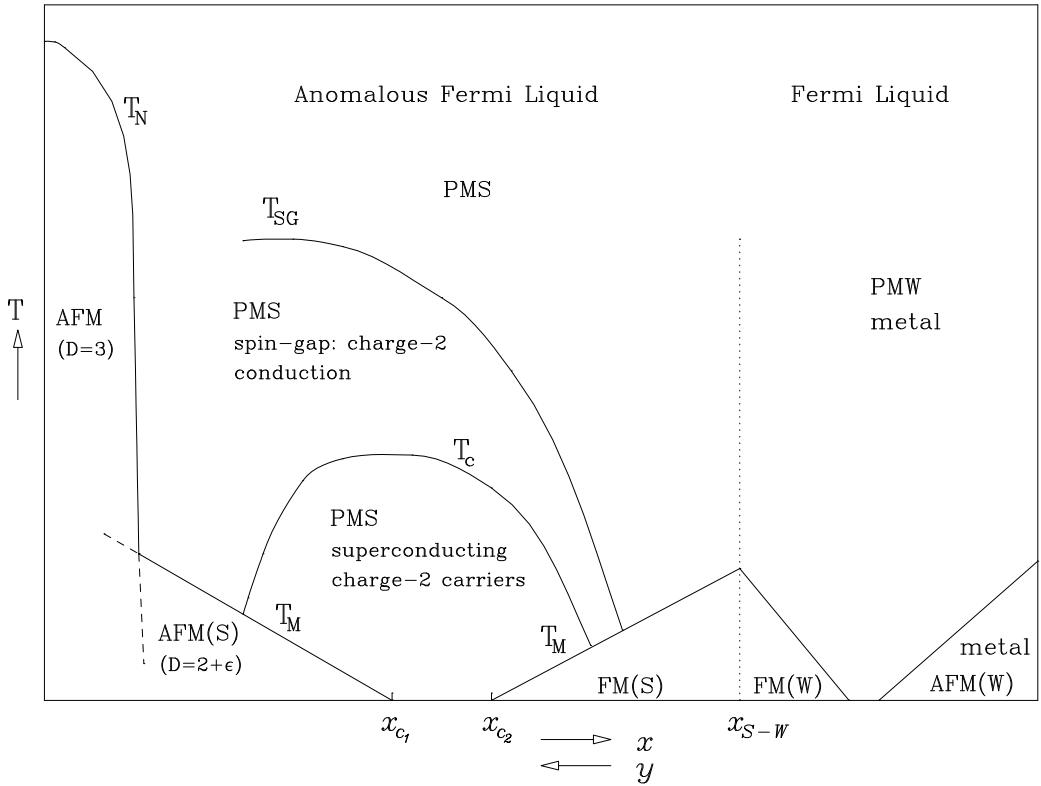,width=0.7\linewidth,angle=90}
\end{center}
\caption{Sketch of the predicted phase structure in the $x$-$T$ plane.
The dotted vertical line at $x_{S-W}$ indicates a possible
crossover.}
\label{XT}
\end{figure}

What happens when the temperature is increased from zero?  So far, we
do not have any MF or MC results at non-zero temperature.
However, by the following fairly general arguments we are led to
conjecture Fig.~\ref{XT} as a schematic sketch of the most general
$x$-$T$ phase diagram suggested by our model.

First, let us discuss what happens with the superconducting phase when
the temperature is increased.
Consider a doping fraction $x$ corresponding to a $y$ value such that we
are well inside the PMS phase, with superconductivity at $T=0$.  At a
certain temperature, $T_\mathrm{BE}$, the BE condensation will be undone,
so the superconductivity disappears.
We thus identify $T_\mathrm{BE}$ as the critical temperature $T_{\mathrm c}$.
Above $T_{\mathrm c}$ the fermions are still bound together in (uncondensed)
light bosonic PMS pairs.
One therefore expects ``normal'', charge-2 conduction due to these pairs.
This phase is sometimes called a ``spin-gap'' phase.
As a matter of fact, the existence of such a phase is shared by any model
in which superconductivity is triggered by the Bose-Einstein condensation
of previously formed fermion-fermion bound states.
Experimental evidence for such a mechanism is now available~\cite{PARES}.

Above some spin-gap temperature, $T_{\mathrm
{SG}}$, the PMS pairs will be broken apart by thermal fluctuations.
However, notice that the pair-breaking temperature for the first, isolated
PMS-pair is lower than the temperatures needed to dissociate many of
them, since the constituent fermions have to fill up states of increasing
energy in the Fermi sea. Thus one
expects $T_{\mathrm {SG}}$ to be characteristic of a cross-over rather
than a well-defined phase transition. In this high-temperature phase, the only
possible carriers would be fermions, but as we have already discussed,
their kinetic term vanishes in the mean field approximation.  (This
perfectly insulating behaviour will receive corrections beyond
the mean-field approximation, though.)

In the qualitative behaviour shown in Fig.\ \ref{XT}, we have supposed
that $T_{\mathrm {SG}}$ is always larger than $T_{\mathrm {BE}}$. 
Depending on the variation of the mass and the binding energy of the
bound state with doping, it could happen that the $T_{\mathrm {SG}}$ and
$T_{\mathrm {BE}}$ critical lines join at some point. 
The actual situation regarding the full curves $T_{\mathrm {SG}}$ and
$T_{\mathrm {BE}}$ would have to be investigated numerically in our model.

Further qualitative statements can be derived from the general thermodynamic
argument that raising the temperature has the effect of increasing the
magnetic disorder.
In the phase diagram of our model, Fig.\ \ref{PHASES},
this means that the two PM phases expand in all directions, at the expense
of the various magnetically ordered phases.
One of the consequences could be a rash disappearance of the FM phase,
which might explain why it has not be observed, as we briefly discussed
in Sect.\ \ref{physzeroT}.

Another consequence is the intriguing possibility of re-entrant
superconductivity for certain values of $x$, 
with superconductivity not starting at $0$ K, but restricted to a temperature interval
\hbox{$0<T_{M} < T < T_{\mathrm c}$.}
To see this, consider $x$ slightly smaller than $x_{c_1}$, corresponding
to a point $(y,k)$ in figure \ref{PHASES} inside the AFM(S) phase, very close to the transition to PMS. 
Let us assume the disorder due to the temperature has similar properties 
to the disorder coming from the dynamics ({\em i.e.} at zero temperature). 
Then the PMS phase will be enlarged and absorb the point $(y,k)$, implying
superconductivity provided that we are still below $T_{\mathrm {BE}}$.
Similar re-entrant behaviour is expected for $x$ slightly larger than
$x_{c_2}$.
In order to make quantitative predictions,
an estimate of the mass and the binding energy of the PMS-pairs
would be required.
Some experimental evidence for reentrance has been found ten years
ago \cite{BREWER} in YBaCuO, but it seems to have gone largely unnoticed.
Further experiments are required to resolve this issue.

In fact, our model does not exclude the extreme case in which the
superconducting phase is an island in the $x$-$T$ plane, completely
detached from the $T=0$-axis.
However, this scenario seems hard to realize in real life,
as it would require the $T=0$ evolution curve
(the arrow line in Fig.\ \ref{PHASES}) to pass underneath but very close to
the PMS phase.

\section{Conclusions}

We have formulated and investigated a simple, 
spin-Dirac fermion field theory in 2+1 dimensions
capable of explaining, at least qualitatively, a variety of
experimental properties of the cuprate superconductors and their parent
compounds as a function of doping fraction and temperature.

Our model provides a qualitative understanding of insulating,
AFM behaviour at low doping, of high-$T_{\mathrm c}$ superconductivity through the
Bose-Einstein condensation of spin-disorder-bound charge pairs at
intermediate doping, and of Fermi-liquid behaviour at large doping.
In addition we have formulated several predictions, which may be amenable
to experimental testing. In particular, we recall the possibility of reentrance
(Sect.\ \ref{nonzeroT}) and the statement that superconductivity is unlikely
to occur in materials with $S >1/2$ (Sect.\ \ref{physzeroT}).

The model is constructed as an effective theory, 
using as ingredients only points 2,4 and 7 of the list of important
experimental properties given in Sect.\ \ref{intro}.
As output, the model gives a reasonable explanation for the other
points 1,3,5 and 6.
We feel this is because we have been able to identify
the essential degrees of freedom, and to pursue the consequences of the
relevant symmetries in classifying their possible interactions.

If we succeed in computing, by means of Monte Carlo simulations,
the masses and the binding energies present of the various states
in the model, we will also be able to study crossover from ``BE-like''
to ``BCS-like'' behaviour.
From the point of view of our model, this crossover
occurs when going from a situation ({\em i.e.,} a $(k,y)$ value)
in which $T_{\mathrm {BE}}<T_{\mathrm {SG}}$, to a situation in which
masses and binding energies are such that $T_{\mathrm {SG}}$ is smaller
than a would-be $T_{\mathrm {BE}}$. In the latter case, the quantum
liquid condensation would occur simultaneously with the pair formation.

The answer to these questions, and to many others, will require much more
future work.
However, let us conclude this paper by trying to anticipate our answers to
the questions listed in Ref.\ \cite{PINES} and quoted in Sect.\ \ref{intro}:

\begin{itemize}
{\item {\bf What is the physical origin of the anomalous {\it normal} state?}
A dynamical, antiferromagnetically interacting spin background, {\it
strongly} coupled to fermions (the {\it heavy} fermions of the type
$\left({\mbox{\boldmath $\tau$}}\cdot{\mbox{\boldmath $\phi$}}\right)\psi$
in our model)}.

{\item{\bf What characterizes this anomalous {\it normal} state?}
The presence of heavy, single fermionic charges 
($\left({\mbox{\boldmath $\tau$}}\cdot{\mbox{\boldmath $\phi$}}\right)\psi$) 
and {\it light, bosonic} charge-2 bound states.} 

{\item {\bf What is the mechanism of high-$T_{\mathrm c}$ superconductivity?}
Bose-Einstein condensation of these dynamically formed, {\it stable}
bosonic charge-2 pairs.}

{\item {\bf What is the pairing state?} A bound state of {\it heavy} fermions,
bound by spin-waves in a {\it disordered\/} phase: a {\bf PMS-pair}.}

\end{itemize}

\section*{Acknowledgments}

We are indebted to A. Muramatsu and L.A.~Fern\'andez for very helpful remarks and
stimulating discussions.
We also acknowledge interesting discussions with A.~Cruz, Ph.~de Forcrand,
J.M. De Teresa, J.G.~Esteve, D.~Frenkel, J.~Garcia, M.R. Ibarra, 
R.~Mahendiran, R.~Navarro,  C.~Rillo, A.~ Taranc\'on and especially G.~Sierra.
We thank, the
RTNN collaboration for computing facilities.
This work is financially supported by CICYT (Spain), projects
AEN 96-1670, AEN 96-1674, AEN 97-1680 and 
by Acci\'on Integrada Hispano-Francesa HF1996-0022.

%123456789%123456789%123456789%123456789%123456789%123456789%123456789%1

\hfill


\newpage

\begin{thebibliography}{99}

\bibitem{BM}
J.G. Bednorz and K.A. M\"uller, {\sl Z.Phys.} {\bf B64} (1986) 189.

\bibitem{RANDERIA}
M. Randeria, preprint cond-mat/9710223.

\bibitem{PARES}
D. Mihailovic {\em et al}, cond-mat/9801049; H. Ding
{\em et al}, cond-mat/9712100.

\bibitem{TSUEI}
C.C. Tsuei {\em et al.}, {\sl Science} {\bf 271} (1996) 329; 
M. Sigrist and T.M.Rice, {\sl Rev. Mod. Phys.} {\bf 67} (1995) 503; 
D.J. Van Harlingen, {\sl Rev. Mod. Phys.} {\bf 67} (1995) 515.

\bibitem{PINES}
D. Pines, preprint cond-mat/9704102.

\bibitem{haldane}
F.D.M. Haldane, {\sl Phys. Lett.} {\bf A93} (1983) 464; {\sl Phys. Rev. Lett.}
{\bf 50} (1983) 1153.

\bibitem{CHAKRA}
S. Chakravarty, B.I. Halperin and D.R. Nelson,
{\sl Phys. Rev. Lett.} {\bf 60}  (1988) 1507;
{\sl Phys. Rev.} {\bf B39} (1989) 2344.

\bibitem{HASENFRATZ}
P. Hasenfratz and H. Leutwyler, 
{\sl Nucl. Phys.} {\bf B343} (1990) 241;
P. Hasenfratz and F. Niedermayer,
{\sl Phys. Lett.} {\bf B268} (1991) 231;
P. Hasenfratz and F. Niedermayer,
{\sl Z. Phys.} {\bf B92}  (1993) 91.

\bibitem{ENDOH} 
Y. Endoh {\em et al.}
{\sl Phys. Rev.} {\bf B37}  (1988) 7443.

\bibitem{DING}
H.Q. Ding and M.S. Makivic
{\sl Phys. Rev. Lett.} {\bf 64}  (1990) 1449.
U.-J. Wiese and H.-P. Ying, {\sl Z. Phys.}\  {\bf B93} (1994) 147.

\bibitem{LETTER}
J.L. Alonso, Ph. Boucaud, V. Mart\'{\i}n-Mayor and A.J. van der Sijs,
preprint cond-mat/9706022, to be published in {\sl Europhys. Lett.}

\bibitem{HEPLAT}
J.L. Alonso, Ph. Boucaud, V. Mart\'{\i}n-Mayor and A.J. van der Sijs,
{\sl Nucl. Phys.} {\bf B (Proc. Suppl.) 63} (1998) 658.

\bibitem{MURAMATSU1}
C. K\"ubert and A. Muramatsu,
{\sl Europhys. Lett.} {\bf 30}  (1995) 481.

\bibitem{ZHANG}
F.C. Zhang and T.M. Rice, {\sl Phys. Rev.} {\bf B37} (1988) 3759.


\bibitem{PARISI}
G. Parisi, {\sl Nucl. Phys.} {\bf B205} (1982) 337;
E. Abdalla, {\sl Phys. Rev.} {\bf D41} (1990) 571.

\bibitem{CHUBUKOV}
A similar result has been found by A.V. Chubukov and D.K. Morr,
preprint cond-mat/9701196.

\bibitem{MARSHALL}
D.S. Marshall {\em et al.}, {\sl Phys. Rev. Lett.} {\bf 76} (1996) 4841; 
{\sl Science} {\bf 273} (1996) 325.

\bibitem{NORMAN}
N.R. Norman {\em et al.}, preprint cond-mat/9710163;
H. Ding {\em et al.}, {\sl Phys. Rev. Lett.} {\bf 78} (1997) 2628.

\bibitem{WIESE97}
B.B. Beard, R.J. Birgeneau, M. Greven and U.-J. Wiese,
preprint cond-mat/9709110.

\bibitem{SCHRIEFFER}
J.R. Schrieffer, X.G. Wen and S.C. Zhang,
{\sl Phys. Rev. Lett.} {\bf 60}  (1988) 944; {\sl Phys. Rev.} {\bf B39}  (1989)
11663.

\bibitem{ANDSCHRI}
P.W. Anderson and J.R. Schrieffer,
{\sl Physics Today}, {\it June  1991}, 54.

\bibitem{BIRGENEAU}
R.J. Birgeneau, {\sl Am. J. Phys.} {\bf 58 (1)}   {\it January 1990}, 28.

\bibitem{AHARONY}
A. Aharony {\em et al.}, {\sl Phys. Rev. Lett.} {\bf 60} (1988) 1330.


\bibitem{AFFLECK}
I. Affleck and J.B. Marston,
{\sl Phys. Rev.} {\bf B37} (1988) 3774.

\bibitem{CONTINUUM}
K. Wilson, in {\it New Phenomena in Sub-Nuclear Physics} (Erice, 1975).
L.H. Karsten and J. Smit, {\sl Nucl. Phys.} {\bf B183} (1981) 103;
H.B. Nielsen and Ninomiya, {\sl Nucl. Phys.} {\bf B185} (1981) 20 and 
{\bf B193} (1981) 173,
Erratum, {\sl Nucl. Phys.} {\bf B195}  (1982) 541.

\bibitem{DE}
A.K. De and J. Jers\'ak, in {\em Heavy flavours}, ed. A. Buras
and M. Lindner (World Scientific Singapore, 1992).

\bibitem{APPELQUIST}
T.W. Appelquist, M. Bowick, D. Karabali and L.C.R. Wijewardhana, 
{\sl Phys. Rev.} {\bf D33}  (1986) 3704.

\bibitem{KARSCH}
For a review see F. Karsch in {\em Quark Gluon Plasma},
(World Scientific Singapore, 1990), edited by R.C. Hwa.

\bibitem{shigemitsubocketal}
See, {\em eg}, J. Shigemitsu,
{\sl Nucl. Phys.}\  {\bf B (Proc.\ Suppl.) 20} 515 (1991);
W. Bock, {\em et al.}, {\sl Nucl. Phys.} {\bf B344}, 207 (1990).

\bibitem{MURAMATSU2}
We are indebted to A. Muramatsu for emphasizing this point.


\bibitem{drouffezuber}
J.-M. Drouffe and J.-B. Zuber,
{\sl Phys.\ Rep.\ } {\bf 102} (1983) 1.

\bibitem{plzar}
J.L. Alonso, Ph.\ Boucaud, F. Lesmes and E. Rivas, 
{\sl Phys.\ Lett.\ } {\bf B329}  (1994) 75.
 
\bibitem{class}
J.L. Alonso, Ph.\ Boucaud, F. Lesmes and A.J. van der Sijs, 
{\sl Nucl.\ Phys.\ } {\bf B457} (1995) 175;
{\bf 472} (1996) 738  (E).

\bibitem{HMC}
R.T. Scalettar, D.J. Scalapino and R.L. Sugar, 
{\sl Phys. Rev.} {\bf B34} (1986) 7911;
S. Duane, A.D. Kennedy, B.J. Pendleton and D. Roweth,
{\sl Phys. Lett.} {\bf B195} (1987) 216.

\bibitem{LIE}
G. Batrouni {\em et al.},
{\sl Phys. Rev.} {\bf D32} (1986) 2736;
S. Gottlieb, W. Liu, D. Toussaint and  R.L. Sugar,
{\sl Phys. Rev.} {\bf D35} (1987) 2531. 

\bibitem{HMCBOOK}
I. Montvay and G. M\"unster,
{\em Quantum Fields on a Lattice} (Cambridge University Press, 1994);
H.J. Rothe,
{\em Lattice Gauge Theories -- An Introduction\/} (World Scientific, 1992).

\bibitem{GOLDSTEIN}
H. Goldstein, {\em Classical Mechanics} (Addison-Wesley, 1959). 

\bibitem{O3}
H. G. Ballesteros, L. A. Fern\'andez, V. Mart\'{\i}n-Mayor and 
A. Mu\~noz Sudupe, {\sl Phys. Lett.} {\bf B387} (1996) 125.

\bibitem{WOLFF}
U. Wolff, {\sl Phys. Rev. Lett.} {\bf 62} (1989) 3834.

\bibitem{NOCLUSTER}
J. L. Alonso {\em et al.}, {\sl Phys. Rev.} {\bf B53} (1996) 2537;
H. G. Ballesteros, L. A. Fern\'andez, V. Mart\'{\i}n-Mayor and 
A. Mu\~noz Sudupe, {\sl Phys. Lett.} {\bf B378} (1996) 207.

\bibitem{GUPTA}
R. Gupta {\em et al.}, {\sl Phys. Rev.} {\bf D40} (1989) 2072.

\bibitem{FS} 
M. Falcioni, E. Marinari, M. L. Paciello, G. Parisi and B. Taglienti,
{\sl Phys. Lett.} {\bf 108} (1982) 331;
A. M. Ferrenberg and R. H. Swendsen, 
{\sl Phys. Rev. Lett.} {\bf 61} (1988) 2635.


\bibitem{STEPHANOV}
M.A. Stephanov and M.M. Tsypin,
{\sl Phys. Lett.} {\bf B236} (1990) 344;
M.A. Stephanov, {\sl Phys. Lett.} {\bf B266} (1991) 447.

\bibitem{SMITz2}
J. Smit, 
{\sl Nucl.\ Phys.\ }{\bf B (Proc.\ Suppl.) \bf 9}  (1989) 579;
M.F.L. Golterman, D.N. Petcher and J. Smit,
{\sl Nucl.\ Phys.\ }{\bf B370} (1992) 51.

\bibitem{SHRAIMAN}
B. I. Shraiman and E. D. Siggia,
{\sl Phys. Rev. Lett.} {\bf 62} (1989) 1564.

\bibitem{MORITOMO}
Y.Moritomo {\em et al.}, {\sl Phys. Rev.} {\bf B51} (1995) 3297

\bibitem{NICKEL} 
G. Blumberg, M.V. Kleain and S. W. Cheong,
{\sl Phys. Rev. Lett.} {\bf 80} (1998) 564; 
J. M. Tranquada, cond-mat/9802043.


\bibitem{BREWER}
J.H. Brewer {\em et al.}, {\sl Phys. Rev. Lett.} {\bf 60} (1988) 1073.

\end{thebibliography}
\end{document}